\documentclass[traditabstract]{aa} 

\usepackage{graphicx}
\usepackage{txfonts}
\usepackage{natbib}
\usepackage[colorlinks=true,urlcolor=dblue,citecolor=dblue,
            linkcolor=dblue]{hyperref}

  \def\H{{\rm\langle H\rangle}}

\begin{document}

\definecolor{GreenYellow}  {cmyk}{0.15,0,0.69,0}
\definecolor{Yellow}{cmyk}{0,0,1,0}
\definecolor{Goldenrod}{cmyk}{0,0.10,0.84,0}
\definecolor{Dandelion}{cmyk}{0,0.29,0.84,0}
\definecolor{Apricot}  {cmyk}{0,0.32,0.52,0}
\definecolor{Peach}    {cmyk}{0,0.50,0.70,0}
\definecolor{Melon}    {cmyk}{0,0.46,0.50,0}
\definecolor{YellowOrange}  {cmyk}{0,0.42,1,0}
\definecolor{Orange}   {cmyk}{0,0.61,0.87,0}
\definecolor{BurntOrange}   {cmyk}{0,0.51,1,0}
\definecolor{Bittersweet}   {cmyk}{0,0.75,1,0.24}
\definecolor{RedOrange}{cmyk}{0,0.77,0.87,0}
\definecolor{Mahogany} {cmyk}{0,0.85,0.87,0.35}
\definecolor{Maroon}   {cmyk}{0,0.87,0.68,0.32}
\definecolor{BrickRed} {cmyk}{0,0.89,0.94,0.28}
\definecolor{Red} {cmyk}{0,1,1,0}
\definecolor{OrangeRed}{cmyk}{0,1,0.50,0}
\definecolor{RubineRed}{cmyk}{0,1,0.13,0}
\definecolor{WildStrawberry}{cmyk}{0,0.96,0.39,0}
\definecolor{Salmon}   {cmyk}{0,0.53,0.38,0}
\definecolor{CarnationPink} {cmyk}{0,0.63,0,0}
\definecolor{Magenta}  {cmyk}{0,1,0,0}
\definecolor{VioletRed}{cmyk}{0,0.81,0,0}
\definecolor{Rhodamine}{cmyk}{0,0.82,0,0}
\definecolor{Mulberry} {cmyk}{0.34,0.90,0,0.02}
\definecolor{RedViolet}{cmyk}{0.07,0.90,0,0.34}
\definecolor{Fuchsia}{cmyk}{0.47,0.91,0,0.08}
\definecolor{Lavender} {cmyk}{0,0.48,0,0}
\definecolor{Thistle}{cmyk}{0.12,0.59,0,0}
\definecolor{Orchid}{cmyk}{0.32,0.64,0,0}
\definecolor{DarkOrchid}{cmyk}{0.40,0.80,0.20,0}
\definecolor{Purple}{cmyk}{0.45,0.86,0,0}
\definecolor{Plum}{cmyk}{0.50,1,0,0}
\definecolor{Violet} {cmyk}{0.79,0.88,0,0}
\definecolor{RoyalPurple} {cmyk}{0.75,0.90,0,0}
\definecolor{BlueViolet}{cmyk}{0.86,0.91,0,0.04}
\definecolor{Periwinkle}{cmyk}{0.57,0.55,0,0}
\definecolor{CadetBlue}{cmyk}{0.62,0.57,0.23,0}
\definecolor{CornflowerBlue}{cmyk}{0.65,0.13,0,0}
\definecolor{MidnightBlue}{cmyk}{0.98,0.13,0,0.43}
\definecolor{NavyBlue} {cmyk}{0.94,0.54,0,0}
\definecolor{RoyalBlue}{cmyk}{1,0.50,0,0}
\definecolor{Blue}{cmyk}{1,1,0,0}
\definecolor{Cerulean} {cmyk}{0.94,0.11,0,0}
\definecolor{Cyan}{cmyk}{1,0,0,0}
\definecolor{ProcessBlue} {cmyk}{0.96,0,0,0}
\definecolor{SkyBlue}{cmyk}{0.62,0,0.12,0}
\definecolor{Turquoise}{cmyk}{0.85,0,0.20,0}
\definecolor{TealBlue} {cmyk}{0.86,0,0.34,0.02}
\definecolor{Aquamarine}{cmyk}{0.82,0,0.30,0}
\definecolor{BlueGreen}{cmyk}{0.85,0,0.33,0}
\definecolor{Emerald}{cmyk}{1,0,0.50,0}
\definecolor{JungleGreen} {cmyk}{0.99,0,0.52,0}
\definecolor{SeaGreen} {cmyk}{0.69,0,0.50,0}
\definecolor{Green}{cmyk}{1,0,1,0}
\definecolor{ForestGreen} {cmyk}{0.91,0,0.88,0.12}
\definecolor{PineGreen}{cmyk}{0.92,0,0.59,0.25}
\definecolor{LimeGreen}{cmyk}{0.50,0,1,0}
\definecolor{YellowGreen} {cmyk}{0.44,0,0.74,0}
\definecolor{SpringGreen} {cmyk}{0.26,0,0.76,0}
\definecolor{OliveGreen}{cmyk}{0.64,0,0.95,0.40}
\definecolor{RawSienna}{cmyk}{0,0.72,1,0.45}
\definecolor{Sepia}{cmyk}{0,0.83,1,0.70}
\definecolor{Brown}{cmyk}{0,0.81,1,0.60}
\definecolor{Tan} {cmyk}{0.14,0.42,0.56,0}
\definecolor{Gray}{cmyk}{0,0,0,0.50}
\definecolor{Black}{cmyk}{0,0,0,1}
\definecolor{White}{cmyk}{0,0,0,0}

\definecolor{StaubBraun}{cmyk}{0,0.61,0.87,0.20}
\definecolor{lyellow}{cmyk}{0,0,0.2,0}
\definecolor{lred}{cmyk}{0,0.4,0.4,0}
\definecolor{lgray}{cmyk}{0,0,0,0.15}
\definecolor{lgreen}{cmyk}{0.3,0,0.8,0}
\definecolor{lblue}{cmyk}{0.4,0,0,0}
\definecolor{dgreen}{rgb}{0.05,0.5,0.1}
\definecolor{dred}{rgb}{0.85,0.0,0.0}
\definecolor{dblue}{rgb}{0.0,0.0,0.6}
\definecolor{brown}{rgb}{0.6,0.1,0.1}

\title{Modelling mid-infrared molecular emission lines from T\,Tauri stars}
   
   \author{P.~Woitke\inst{1,2}, 
           M.~Min\inst{3},
           W.-F.~Thi\inst{4},
           C.~Roberts\inst{1},
           A.~Carmona\inst{5},
           I.~Kamp\inst{6},
           F.~M{\'e}nard\inst{7},
           C.~Pinte\inst{7}
   }
   
   %\titlerunning{Modelling infrared molecular emission lines
   %              from protoplanetary discs}
   \authorrunning{P.~Woitke et al.}

   \institute{ %1 Woitke
             SUPA School of Physics \& Astronomy, University of St Andrews,
             North Haugh, KY16\,9SS, St Andrews, UK
          \and %2 Woitke
             Centre for Exoplanet Science, University of St Andrews, 
             St Andrews, UK
          \and %3 Min
             Astronomical Institute ``Anton Pannekoek'', University of 
             Amsterdam, PO Box 94249, 1090 GE Amsterdam, The Netherlands
          \and %4 Thi
             Max Planck Institute for Extraterrestrial Physics,
             Giessenbachstrasse, 85741 Garching, Germany
          \and %5 Carmona
             Universit{\'e} de Toulouse, UPS-OMP, IRAP, 14 avenue
             E.~Belin, Toulouse, 31400, France
          \and %6 Kamp
             Kapteyn Astronomical Institute, Postbus 800,
             University of Groningen, 9700 AV Groningen, The Netherlands
          \and %7 Menard, Pinte
             Univ.~Grenoble Alpes, CNRS, IPAG, F-38000 Grenoble, France
             }

   \date{Received\ \ June\,28, 2017;\ \ accepted\ \ July\,8, 2018}

   \abstract{We introduce a new modelling framework including the {\sl
       Fast Line Tracer} ({\sc FLiTs}) to simulate infrared line
     emission spectra from protoplanetary discs. This paper focusses
     on the mid-IR spectral region between 9.7\,$\mu$m to 40\,$\mu$m
     for T\,Tauri stars.  The generated spectra contain several tens
     of thousands of molecular emission lines of H$_2$O, OH, CO,
     CO$_2$, HCN, C$_2$H$_2$, H$_2$, and a few other molecules, as well
     as the forbidden atomic emission lines of S\,I, S\,II, S\,III,
     Si\,II, Fe\,II, Ne\,II, Ne\,III, Ar\,II, and Ar\,III.  In contrast
     to previously published works, we do not treat the abundances of
     the molecules nor the temperature in the disc as free parameters,
     but use the complex results of detailed 2D {\sc ProDiMo} disc
     models concerning gas and dust temperature structure, and
     molecular concentrations. {\sc FLiTs} computes the line emission
     spectra by ray tracing in an efficient, fast and reliable
     way. The results are broadly consistent with $R\!=\!600$
     Spitzer/IRS observational data of T\,Tauri stars concerning line
     strengths, colour, and line ratios.  In order to achieve that
     agreement, however, we need to assume either a high gas/dust mass
     ratio of order 1000, or the presence of illuminated disc walls at
     distances of a few~au, for example due to disc-planet
     interactions. These walls are irradiated and heated by the star
     which causes the molecules to emit strongly in the mid-IR. The
     molecules in the walls cannot be photodissociated easily by UV
     because of the large densities in the walls favouring their
     re-formation.  Most observable molecular emission lines are found
     to be optically thick. An abundance analysis is hence not
     straightforward, and the results of simple slab models concerning
     molecular column densities can be misleading. We find that the
     difference between gas and dust temperatures in the disc surface
     is important for the line formation. The mid-IR emission features
     of different molecules probe the gas temperature at different
     depths in the disc, along the following sequence:\ \ OH
     (highest)\ \ --\ \ CO\ \ --\ \ H$_2$O and
     CO$_2$\ \ --\ \ HCN\ \ --\ \ C$_2$H$_2$ (deepest),\ \ just where
     these molecules start to become abundant. We briefly discuss the
     effects of C/O ratio and choice of chemical rate network on these
     results. Our analysis offers new ways to infer the chemical and
     temperature structure of T\,Tauri discs from future James Webb
     Space Telescope (JWST)/MIRI observations, and to possibly detect
     secondary illuminated disc walls based on their specific mid-IR
     molecular signature.}

   \maketitle

%============================================================================
\section{Introduction}
%============================================================================

The {\sc Spitzer Space Telescope} was the first astronomical
instrument to detect water and simple organic molecules in
protoplanetary discs at mid-IR wavelengths. \citet{Lahuis2006}
reported on absorption bands of CO$_2$, HCN and C$_2$H$_2$ in
the spectrum of the embedded low-mass young stellar object IRS~46,
attributed to molecules in the line of sight within a distance of few
au from this embedded star. Further early detections from Spitzer
concerned the mid-IR H$_2$ lines \citep{Nomura2005} and the
[Ne\,II]\,12.81\,$\mu$m line \citep{Pascucci2007,Gudel2010}, the
latter showing slightly blue-shifted profiles in follow-up VISIR
high-spectral resolution observations by \citet{Pascucci2009b},
interpreted as evidence for a slow disc wind.  Other forbidden lines
were detected by \citet{Baldovin-Saavedra2011}, in particular
[Fe\,II] lines, whereas [Ne\,III] and [S\,III] were not found.

The detection of mid-IR H$_2$O lines by Spitzer/IRS was announced
simultaneously by \citet{Carr2008} for AA\,Tau and by
\citet{Salyk2008} for DR\,Tau and AS\,205. A rich variety of H$_2$O
and OH molecular line emission features were observed in addition to
some of the already detected molecules and ions listed above.
%\citet{Sargent2014} detected H$_2$O in emission around $6.6\,\mu$m 
% in 7 objects.
\citet{Pontoppidan2010} re-reduced archival Spitzer/IRS spectra and
established that the majority of T\,Tauri stars show water emission
features in the $10-36\,\mu$m spectral range, originating from the
$0.1-2$\,au radial disc region with characteristic emission
temperatures of about $300-1000$\,K.  This analysis was generally
confirmed later by \citet{Salyk2011} and \citet{Carr2011}.
%, however some discrepancies with respect to the results of
%Pontoppidan et al.\ have been noted in terms of the derived column
%densities, possibly because some of these lines are blended, and
%simple LTE slab models have been designed in different ways for the
%interpretation.

The analysis of the emission lines of TW\,Hya by \citet{Najita2010}
revealed strong OH lines, in particular at longer wavelengths
$\sim\!20-36\,\mu$m, and high-excitation HI lines such as\linebreak
HI\,(9-7) at 11.31\,$\mu$m and HI\,(7-6) at 12.37\,$\mu$m.
\citet{Rigliaco2015} carefully re-reduced a large number of
$R\!=\!600$ Spitzer/IRS spectra of T\,Tauri stars. These authors
concluded that the hydrogen lines do not trace the gas in the disc,
but rather the gas accreting onto the star in the same way as other
hydrogen recombination lines do at shorter wavelengths.

\citet{Pascucci2009,Pascucci2013} reported on variations of the
C$_2$H$_2$ and HCN line strengths with stellar spectral type.
\citet{Najita2013} related the HCN/water ratio to the carbon to oxygen
(C/O) element abundance ratio. One current interpretation is that the
C/O ratio might be considerably larger than solar in the
planet-forming region of protoplanetary discs, varying with stellar
type and varying between individual objects
\citep[e.g.][]{Du2015,Kama2016,Walsh2015}. However, high-resolution
optical spectra of Herbig Ae stars show that the C/O ratio of the gas
that is accreted onto the stars always has the same, solar-like value
\citep{Folsom2012,Kama2015}. An interesting observation is the
detection of the C$_2$H$_2$ emission feature at 13.7\,$\mu$m in some
T\,Tauri stars because at solar C/O this molecule should only be
abundant in deeper layers that are already optically thick in the
continuum \citep{Agundez2008,Walsh2015}. However, this conclusion
depends on our possibly incomplete understanding of the warm chemistry
in discs, and mixing processes could increase the concentrations of
such molecules in upper, observable disc layers \citep{Semenov2011}.

Herbig Ae/Be stars show much lower detection rates of mid-IR molecular
emission lines with Spitzer/IRS in comparison to T\,Tauri stars, in the
form of tentative detections of H$_2$O and OH beyond 20\,$\mu$m
\citep{Pontoppidan2010} only in a few cases. It has been suggested
that the missing water lines could be caused by some specific chemical
or excitation effects in the warmer Herbig\,Ae discs
\citep{Meijerink2009,Fedele2011} or by growing inner cavities
\citep{Banzatti2017}.  However, these lines are also buried in a
stronger continuum. \citet{Antonellini2016} have {argued} that
common data reduction techniques, such as pattern noise and
de-fringing, {are likely to produce higher noise levels when the
  continuum is stronger. Therefore, the Spitzer observations 
might have been simply not deep enough to detect a similar level of 
mid-IR molecular emission from Herbig Ae/Be stars.}

Simple slab models have been used in most cases for the analysis of
these data in terms of molecular column densities and disc
temperatures. In these slab models, a single temperature and fixed
molecular concentrations are considered, and the excitation of the
molecules is approximated in local thermodynamic equilibrium
(LTE). The integration along the line of sight can be solved
analytically in this case, hence these models are actually
single-point (0D) models, and the results only depend on the
temperature and molecular column densities assumed. The solid angle of
the emitting region is adjusted later to match the observed strength
of the molecular emission features. However, these slab models come in
a number of variants. The lines of different molecules from the same
disc are often fitted with different physical slab-parameters. Also
the way in which the dust continuum is treated in these models differs
considerably among different authors. In the near and mid-IR, dust
opacity effects are likely to be very important as large amounts of
molecules are expected to be present below the $\tau_{\rm dust}=1$
surface where they do not produce any observable signatures.

The concept of LTE-slab models can be generalised to non-LTE, where
again a single temperature and a single gas density is considered,
for example the RADEX code \citep{vanderTak2007}.  The molecular level
populations are computed at given molecular column density using the
escape-probability formalism, and background radiation fields in
the form $J_\nu\!=\!W B_\nu(T_{\rm rad})$ can be included, an effect
called radiative pumping. Like in LTE slab models, the integration
along the line of sight can be solved analytically, but in addition to
the temperature $T$ and the molecular column density, the non-LTE slab
model results also depend on volume density, dilution factor $W$, and
radiative temperature $T_{\rm rad}$.

Series of slab models can be used to better represent the changing
physical conditions with radius along the disc surface, either
assuming LTE or non-LTE, in particular for CO fundamental
ro-vibrational emission
\citep{Blake2004,Thi2005,Thi2005b,Brittain2009,
  Ilee2013,Ilee2014,Carmona2017}. These models usually use power laws
for molecular column densities and temperature as a function of radius.
Such 1D slab models are then integrated over the radius to compute
the total line emission fluxes from the disc.

\citet{Bruderer2015} compiled a non-LTE model for the HCN molecule
from the literature for ro-vibrational levels.  A combination of LTE
and non-LTE slab models was performed, as well as a full 2D disc model
of the T\,Tauri star AS\,205\,(N), which is well-known for its
exceptionally strong molecular emission lines in the IR and far-IR
spectral regions \citep{Salyk2011, Fedele2013}. However, Bruderer et
al.\ did not use their own consistent results for gas temperature and
molecular concentrations, but have chosen to assume $T_{\rm
  gas}\!=\!T_{\rm dust}$ and to introduce parameterised
jump-abundances to avoid the complexity of
heating/cooling balance and kinetic chemistry in discs for the
interpretation of the spectra. These authors found a critical density for the
population of the upper vibrational state of the HCN 14\,$\mu$m
feature of order $\sim\!10^{12}\rm\,cm^{-3}$, and concluded that
non-LTE effects can increase the mid-IR line fluxes by up to a factor
of about three. Concerning the hot 3\,$\mu$m band of HCN, non-LTE
effects can also cause extended emission, leading to more centrally
peaked lines.

Similar investigations have recently been carried out by
\citet{Bosman2017} for the CO$_2$ molecule in AS\,205\,(N). The
authors found a critical density for the population of the upper
vibrational level of the 15\,$\mu$m CO$_2$ emission feature of
$\sim\!10^{12}\rm\,cm^{-3}$ and arrived at similar conclusions about
the importance of non-LTE effects. They also present first predictions
for the IR spectrum of the $^{13}$CO$_2$ molecule in consideration of
JWST. However, the assumption of $T_{\rm gas}\!=\!T_{\rm dust}$ and
the parameterised jump-abundances are likely to produce new
uncertainties. In particular, the parameterised abundance causes the
molecules to be present already at very high altitudes, where
densities are low, {radiation fields are strong, and hence
non-LTE effects are likely to be important, whereas in our disc models
molecules such as CO$_2$ and HCN are only abundant in deeper layers
in which densities are larger and non-LTE effects are less important.}

In new investigations, based on full 2D {\sc ProDiMo} thermochemical
disc models, \citet{Antonellini2015} have shown that the mid-IR
water lines are excited very close to LTE, originate from different
radii with different temperatures, and are affected by stellar
luminosity and various disc properties, such as dust opacities,
stellar UV irradiation, dust/gas ratio, dust settling, and disc inner
radius. The emission spectra calculated by \citet{Antonellini2015} use
the vertical escape probability technique (see Appendix~\ref{AppA}),
which is strictly valid only for face-on discs.

%============================================================================
\section{Purpose of this paper}
%============================================================================

In this paper, we run full 2D {\sc ProDiMo} thermochemical disc models
to predict the dust and gas temperature structures and molecular
concentrations, and then perform a {global} ray-tracing technique to
simulate the molecular emission lines from protoplanetary discs.  We
introduce {\sc FLiTs}, the fast line tracer, to ray trace the entire
disc for arbitrary inclination angles.  For demonstration, we use this
new modelling platform to simulate the entire line emission spectra of
T\,Tauri discs in the mid-IR wavelength region, where the
observational data contains a wealth of spectroscopically unresolved
emission lines that merge into complicated spectral emission
features. We convolve our high-resolution spectra to a spectral
resolution of $R\!=\!600$ to compare the spectra to a small collection of
Spitzer/IRS spectra of T\,Tauri stars from \citet{Rigliaco2015}.

We do not intend to fit any particular observational data in this
paper, but we want to discuss to what extent our model spectra are
broadly consistent with the observed line strengths and spectral
shapes of the emission features for various molecules. We investigate
which disc properties are responsible for the strength and colour of
line emission, including disc shape and dust opacities. We want to
explore to what extent the chemical concentrations predicted by {\sc
 ProDiMo} are consistent with the Spitzer/IRS data, or whether
certain molecules are possibly underpredicted or overpredicted in
systematic ways. We briefly discuss the underlying uncertainties
in chemical rate networks and the carbon to oxygen ratio.

These are a few first steps towards solving the more important, general
scientific questions connected to infrared line emission spectra from
protoplanetary discs.
\begin{itemize}
\setlength{\itemsep}{1.3pt}
\setlength{\parskip}{0pt}
\setlength{\topsep}{0pt}
\setlength{\parsep}{0pt}
\setlength{\partopsep}{0pt}
\item What is the molecular composition of the warm gas in the
  planet-forming regions of protoplanetary discs?
\item Why do some T\,Tauri stars show strong molecular emission lines
  whereas others do not? Why do Herbig\,Ae stars show lower detection
  rates?
\item What are the principal chemical and physical processes to
   excite the molecular emission lines, and how tightly are these related
   to stellar properties such as UV excess and X-ray luminosities?
\item Can we infer the vertical gas temperature structure in the
  planet-forming region from the observation of IR molecular emission
  lines?
\item Can we use IR molecular emission lines to diagnose disc winds
  and disc shape anomalies such as gaps, vortices, and spiral waves at
  radial distances of a few au?
\item What can we conclude about dust opacities and disc evolution in the
  planet-forming region of protoplanetary discs?
\item What are the predictions for JWST/MIRI in the near and
  SPICA/SMI in the distant future, which new science questions can we
  address, and what are the prospects of discovering new molecules at
  IR wavelengths?
\item Is there evidence for an enrichment of the gas with elements
  outgassing from comets/pebbles that migrate inwards? Can this
  process cause a carbon-rich local environment?
\end{itemize}
Archival (e.g. Spitzer/IRS) and future observational spectra
(e.g.\ JWST/MIRI, in the distant future SPICA/SMI) will contain
valuable information about the physico-chemical state of
protoplanetary discs in the planet-forming regions as probed by mid-IR
line emission, but in order to deduce this information from the data
and to address the questions above, we need to go beyond simple
isothermal slab models in LTE. We require disc models that include a
detailed treatment of the disc structure, dust radiative transfer, gas
heating and cooling, chemistry, and line radiative transfer.

The future JWST/MIRI data will improve by a factor of about five in
spectral resolution and will have a sensitivity significantly higher
than the current Spitzer/IRS data.  However, for JWST/MIRI, we still
have to face the challenge of analysing spectrally unresolved data. To
assist with the correct interpretation and to address special
questions concerning the dynamics of gas and winds, follow-up
observations at high spectral resolution ($R\!\ga\!17000$) are
required.  Such observations have been carried out using ground-based
near-IR and mid-IR spectrographs such as the Very Large Telescope
(VLT)/CRIRES \citep{Thi2013,Carmona2014}, VLT/VISIR
\citep{Pontoppidan2010b,Pascucci2011,Baldovin-Saavedra2012,Sacco2012,Banzatti2014},
GEMINI/TEXES \citep{Salyk2015} or GEMINI/MICHELLE \citep{Herczeg2007}.
The {\sc ProDiMo} models have already been used successfully to
interpret high-resolution near-IR data
\citep[e.g.][]{Bertelsen2014,Bertelsen2016a,Bertelsen2016b}, and the
new {\sc ProDiMo} $+$ {\sc FLiTs} models introduced in this paper are
expected to provide an excellent basis to analyse high-resolution
mid-IR data in the future. However, in this paper, we concentrate on
discussing the spectrally unresolved Spitzer/IRS data and future
prospects for JWST/MIRI.

%\citep{Pontoppidan2010b} water, VISIR
%\citep{Pascucci2011} Ne\,II, VISIR
%\citep{Salyk2015} water, TEXES
%\citep{Herczeg2007} Ne\,II, MICHELLE
%\citep{Banzatti2014} water, VISIR
%\citep{Sacco2012} Ne\,II, VISIR

%============================================================================
\section{Disc model}
%============================================================================

To simulate the mid-IR emission line spectra from T\,Tauri stars, we
use the radiation thermochemical disc models described by
\citet{Woitke2016}. The models are based on {\sl MCFOST}
\citep{Pinte2006} and {\sl MCMax} \citep{Min2009} Monte-Carlo
radiative transfer coupled to disc chemistry, gas heating and cooling
balance, and non-LTE line formation performed by {\sc ProDiMo}
\citep{Woitke2009,Kamp2010,Thi2011}.  In this paper, we use a generic
model set-up for a T\,Tauri disc as described in \citep{Woitke2016}.  A
protoplanetary disc of mass $0.01\,$M$_\odot$ is considered around a
K7-type young star of mass $M_\star=0.7\,$M$_\odot$ and stellar
luminosity $L_\star=1\,$L$_\odot$, with an age of about $1.6\,$Myrs at
a distance of 140\,pc. The disc is seen under an inclination of
$45^\circ$.  The disc is assumed to start at the dust sublimation
radius of $R_{\rm in}\!=\!0.07$\,au. Further disc shape and dust
opacity parameters of this model are listed in \citep[][see Table~3
  therein]{Woitke2016}.  

In the main model selected for our broad comparison to Spitzer/IRS
observations in this paper, however, we increased the gas/dust mass
ratio (at constant dust mass) from 100 to 1000 to reach the observed
strengths of the various line emission features. This idea was first
introduced by \citet{Meijerink2009}, modelling H$_2$O lines, and then
later also used by \citet{Bruderer2015} and
\citet{Bosman2017}. Gas/dust ratios of order 1000 can readily be
obtained in evolutionary disc models that take into account dust
migration, for example after about 1\,Myrs in the models of
\citet{Birnstiel2012}, as shown by Greenwood et al.\ (in prep.).  The
actual disc gas mass of the main model would therefore actually be
$0.1\,$M$_\odot$.  However, since the mid-IR lines originate well
inside of 10\,au, we only need to apply this modification to those
regions, which does not reflect on the overall disc mass as only about
10\% of the mass resides in those inner regions.  Alternative ideas
for how to increase the mid-IR line strengths (removal of small
grains, e.g.\ by very strong dust settling, disc gaps with subsequent
vertical walls) are briefly discussed in Sect.~\ref{sec:gas_dust}.

\begin{figure*}
\centering
\vspace*{-1mm}
\includegraphics[width=8.3cm,trim=10 4 4 4, clip]{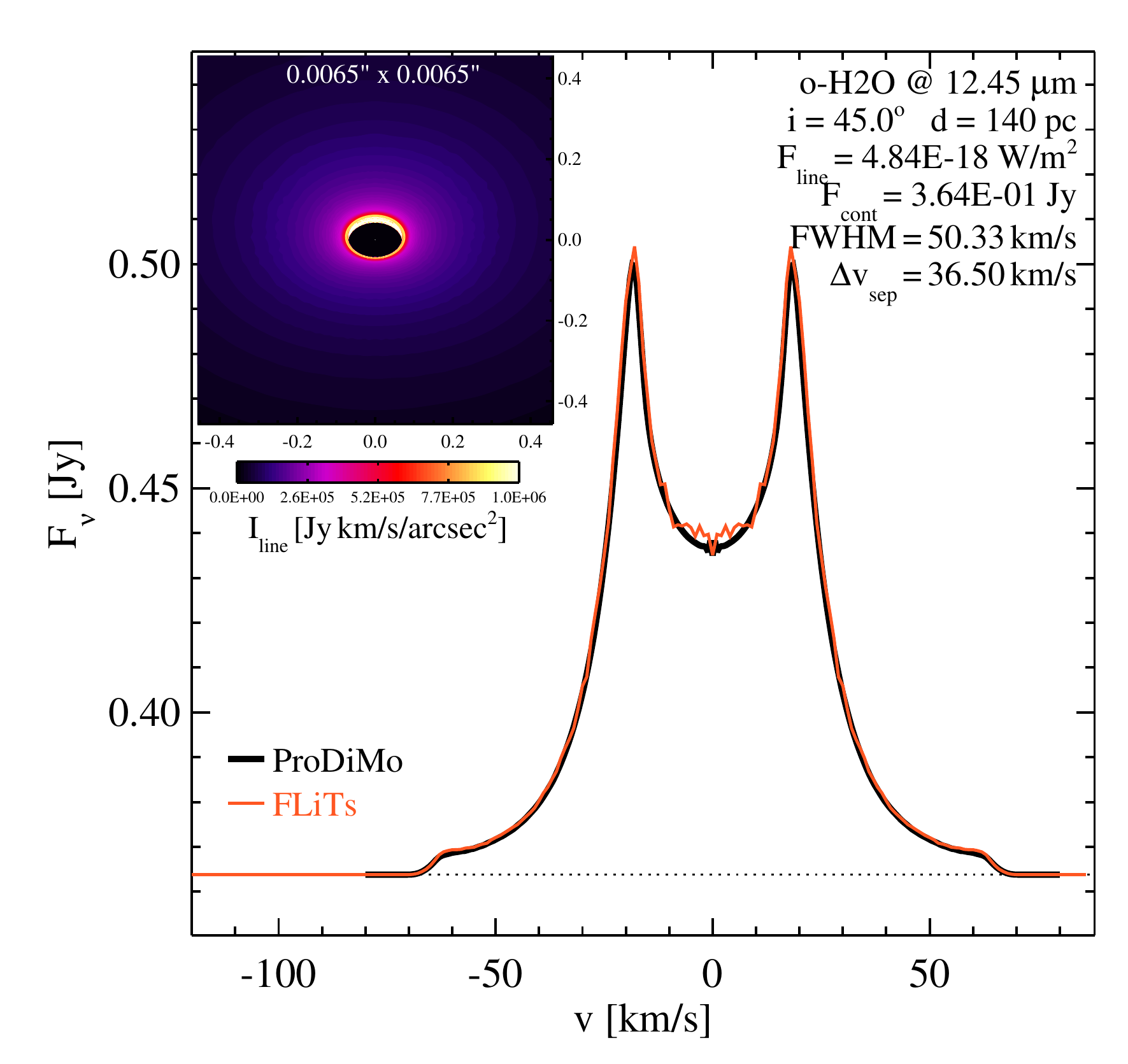}
\includegraphics[width=8.3cm,trim=10 4 4 4, clip]{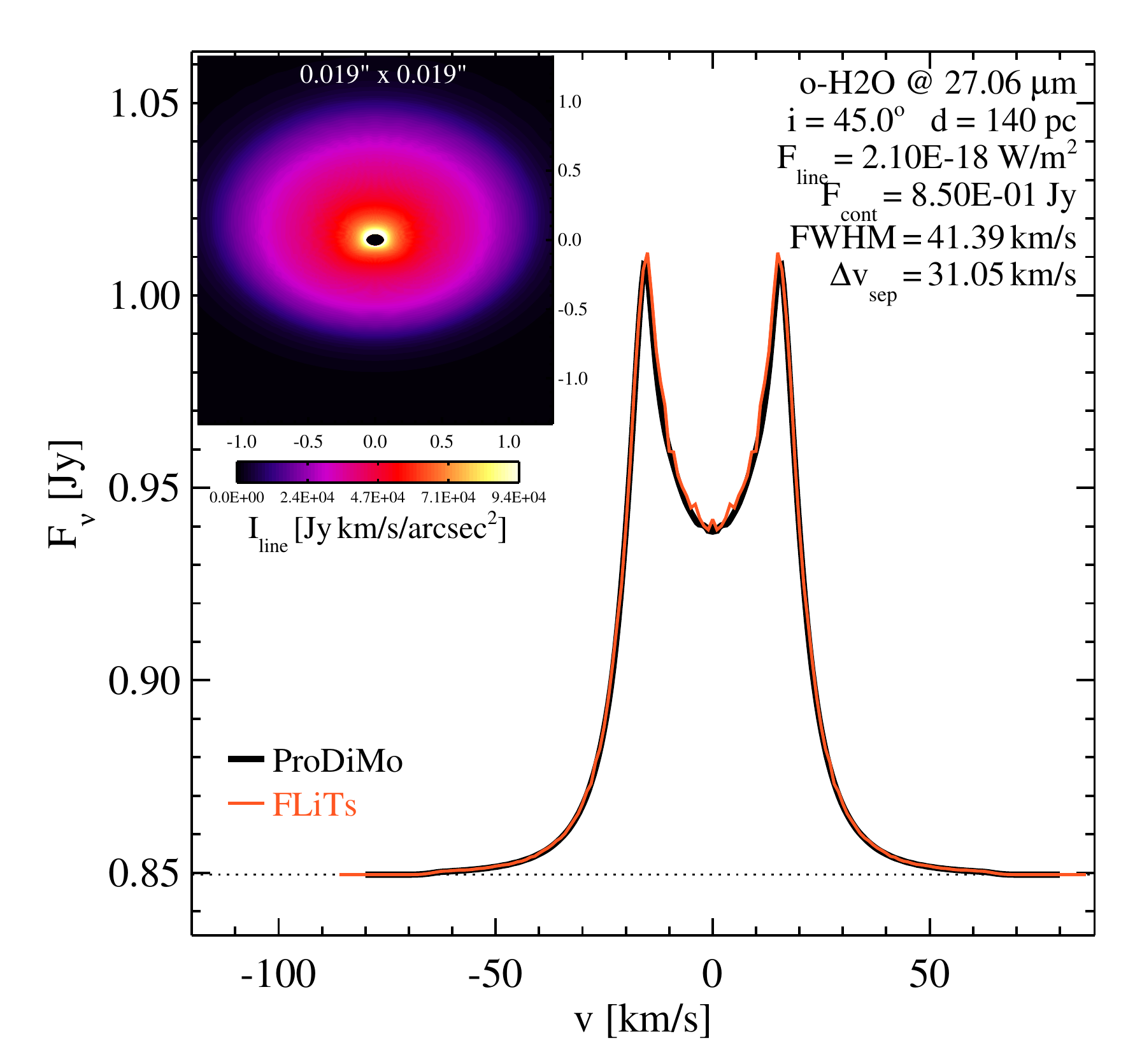}\\[-2mm]
\caption{Computation of high-resolution single line profiles with {\sc
    ProDiMo} (black lines) and {\sc FLiTs} (red lines) based on a
  medium-resolution 90$\times$70 disc model. On the left side we show
  the o-H$_2$O rotational line $\rm 13_{\,6,5}\to 12_{\,4,9}$ at
  12.453\,$\mu$m with an excitation energy of 4213\,K, which is
  emitted partly from the inner rim of the disc, and partly from the
  disc surface up to a radius of about 0.4\,au in this model.  On the
  right side, the o-H$_2$O rotational line $\rm 8_{\,5,4}\to
  8_{\,2,7}$ at 27.059\,$\mu$m is depicted, which has an excitation
  energy of 1806\,K and is mainly emitted from the disc surface up to
  a radius of about 1\,au. The time consumption is about 92\,CPU-sec
  with {\sc ProDiMo} ($251$ velocity-channels, $356\times 144$ rays)
  and 1.2\,CPU-sec with {\sc FLiTs} with enhanced accuracy settings
  (173 channels, 22375 rays), i.e.\ enhanced over {\sc FLiTs}
  standards.}
\label{FLiTs-ProDiMo1}
\vspace*{-1mm}
\end{figure*}
% ProDiMo/DIANA_standard/with_FLiTs10_large_dg1000_LTE
% o-H2O
% 138 ->   84 lam[mic]= 1.245346E+01 Eu[K]= 4.213E+03 Aul[1/s]= 1.160E+00 Ncr[cm^-2]= 3.521E+14     0    0    0,    13    7    6->     0    0    0,    12    4    9
% o-H2O
%  40 ->   28 lam[mic]= 2.705872E+01 Eu[K]= 1.806E+03 Aul[1/s]= 4.545E-02 Ncr[cm^-2]= 9.460E+14     0    0    0,     8    5    4->     0    0    0,     8    2    7

In the first modelling step, we calculated the opacities of the dust
particles for sizes between $0.05\,\mu$m and $3\,$mm \citep[see
  details in][]{Min2016}, considering about 100 dust size bins. A
power-law size distribution was considered with constant dust/gas mass
ratio throughout the disc, however in each column, we first computed
the density-dependent settled size distribution according to
\citet{Dubrulle1995}, before summing up the total dust opacities at
each point in the disc. We then perform a full 2D radiative transfer
either by {\sc MCFOST}, {\sc MCMax}, or {\sc ProDiMo} (can be used
interchangeably) to obtain the dust temperature structure $T_{\rm
  dust}(r,z)$.

The disc model uses the large DIANA chemical standard with 235 gas and
ice species \citep{Kamp2017} and 90 heating and 83 cooling rates to
compute the gas temperature and chemical concentrations at every point
in the disc in local energy balance and kinetic chemical
equilibrium. The level populations of the various atomic and molecular
species are usually calculated in non-LTE in {\sc ProDiMo}. However,
with respect to previous publications, we included a few more
ro-vibrational molecules from the HITRAN database
\citep{Rothman1998,Rothman2013}, approximating their level populations
in LTE for simplicity.  Table~\ref{tab:IRdata} summarises the relevant
spectroscopic data in the mid-IR spectral region. Concerning the
non-LTE atoms and molecules, an escape probability method is used to
compute the level populations \citep{Woitke2009}, based on the
calculated molecular column densities and continuum radiative transfer
results. Thus, the IR-pumping of the level population by the emission
and scattering of dust particles in the disc is fully taken into
account.  All spectral lines considered are automatically included as
additional heating/cooling processes in {\sc ProDiMo}.

We emphasise that we are using the full {\sc ProDiMo} results in this
paper (gas and dust temperatures, settled dust opacities, continuum
mean intensities, chemical concentrations, and level populations) to
compute the mid-IR molecular spectra of class\,II T\,Tauri discs if
not otherwise stated, and this is different from\ \citet{Bruderer2015}
and \citet{Bosman2017}, for example.

The mid-IR line emitting regions in our disc model satisfy the
physical condition of local energy conservation, which we consider as
an advantage of our modelling approach.  For example, the total line
luminosity produced by these disc regions cannot exceed the total
amount of energy that these regions receive from the star, either
directly in the form of X-ray and UV photons triggering various heating
processes, or indirectly via stellar photons that are processed by the
disc to cause a strong local IR radiation field, which then heats the
gas via line absorption. This is all taken into account in our model,
which enables us to discuss, for example, how different disc
geometries and different stellar irradiation properties produce
different line emission spectra (see Sect.~\ref{sec:Rin}). Important
heating and cooling processes in the line emitting regions are further
discussed in Sect.~\ref{sec:details}.

%============================================================================
\section{FLiTs -- the fast line tracer}
%============================================================================
The {\sl Fast Line Tracer} ({\sc FLiTs}) has been developed by M.~Min
to quickly and accurately compute the rich molecular emission line
spectra from discs in the infrared, although it can principally be
applied in all wavelength domains. It is a stand-alone {\sc Fortran-90}
module designed to read the output from {\sc ProDiMo}
and perform the line ray tracing to simulate the observations of
molecular line spectra\footnote{{\sc FLiTs} is available on a
  collaborative basis; please contact M.~Min.}.

The main challenges in this wavelength domain are (i) large numbers of
molecular levels and lines, i.e.\ potentially tens of thousands of levels
per molecule and millions of lines, for example known for H$_2$O and
CH$_4$; (ii) physical overlaps of spectral lines, i.e.\ line photons
emitted by one part of the disc may be absorbed in a different line by
another part of the disc; and (iii) high optical depths in both line
and continuum, which requires full radiative transfer
solutions. Section~\ref{sec:Levels} describes how a careful selection
of molecules, levels, and lines can be made, while maintaining
scientific significance, to bring the computational efforts down to a
manageable level. In this section, we concentrate on the
remaining technical challenges.

\begin{figure}[!t]
\centering
\vspace*{1mm}
\includegraphics[width=9.2cm,trim=15 10 4 20, clip]{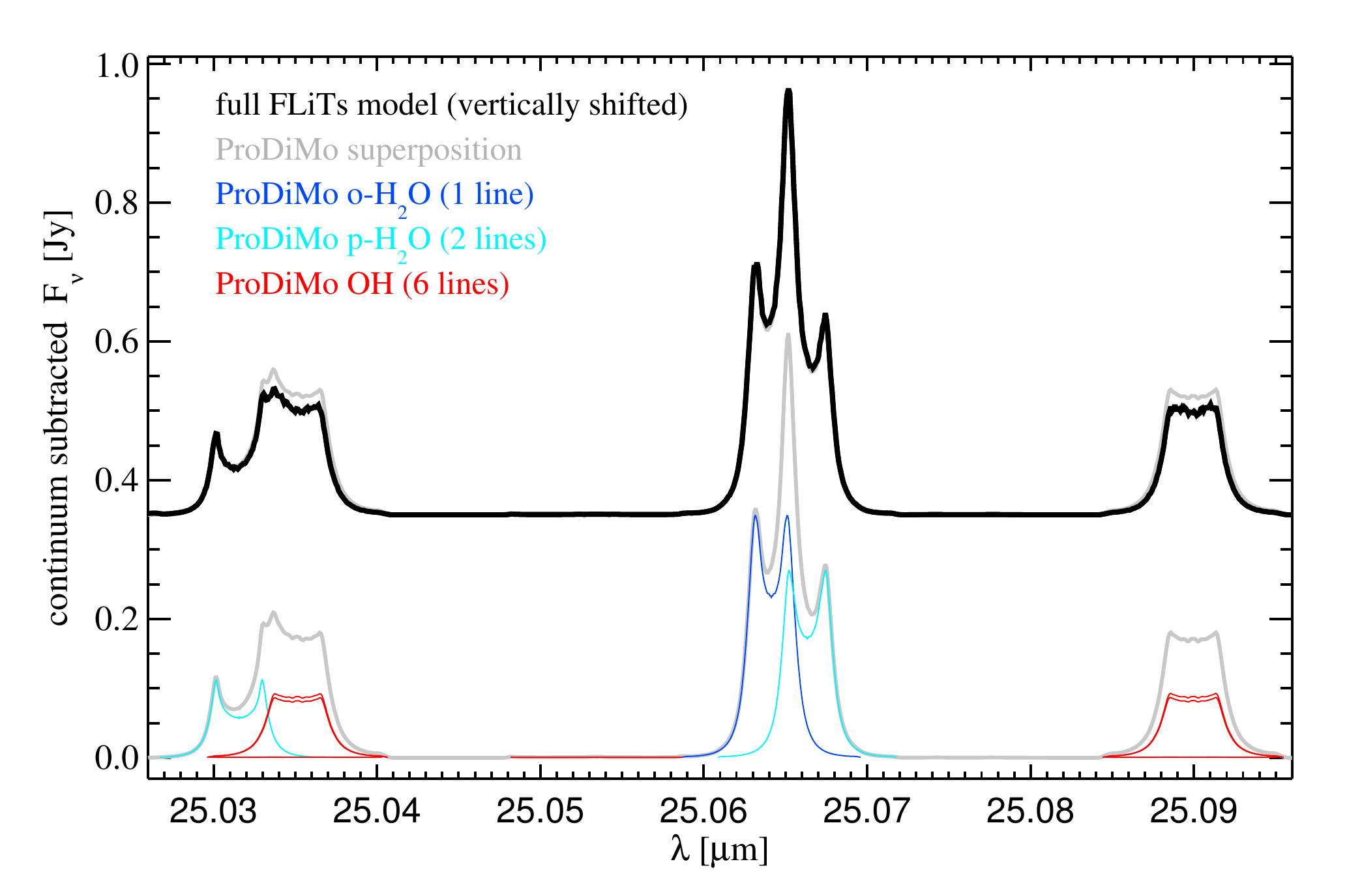}
\vspace*{-4mm}
\caption{Computation of overlapping OH and H$_2$O lines with {\sc FLiTs}
  and {\sc ProDiMo} around 25\,$\mu$m. The central feature shows two 
  about equally strong water lines:
  o-H$_2$O (blue) $9_{7,2}\to 8_{6,3}$ at 25.0641\,$\mu$m and
  p-H$_2$O (cyan) $9_{7,3}\to 8_{6,2}$ at 25.0663\,$\mu$m. 
  Although these two lines overlap spectroscopically, they do not overlap
  physically, and superposition (grey) still works fine in
  comparison to the full {\sc FLiTs} model (black). On the right side, 
  there are 3 individual OH lines (hyper-fine splitting) that 
  physically overlap,
  $X_{3/2},v\!=\!0,J\!=\!12\!\to\!X_{3/2},v\!=\!0,J\!=\!11$ 
  at 25.089950\,$\mu$m (red, strong),
  $X_{3/2},v\!=\!0,J\!=\!11\!\to\!X_{3/2},v\!=\!0,J\!=\!10$ 
  at 25.089946\,$\mu$m (red, strong), and
  $X_{3/2},v\!=\!0,J\!=\!11\!\to\!X_{3/2},v\!=\!0,J\!=\!11$ 
  at 25.089835\,$\mu$m (red, very weak).
  The superposition gives slightly too strong results. 
  On the left side, another case is shown with 1 p-H$_2$O
  line and 3 OH lines.}
\label{FLiTs-ProDiMo2}
\vspace*{-1mm}
\end{figure}

\begin{figure*}[!t]
\vspace*{-3mm}
\centering
\begin{tabular}{cc}
\hspace*{-7mm}
\includegraphics[width=9.7cm,trim=10 11 4 4, clip]{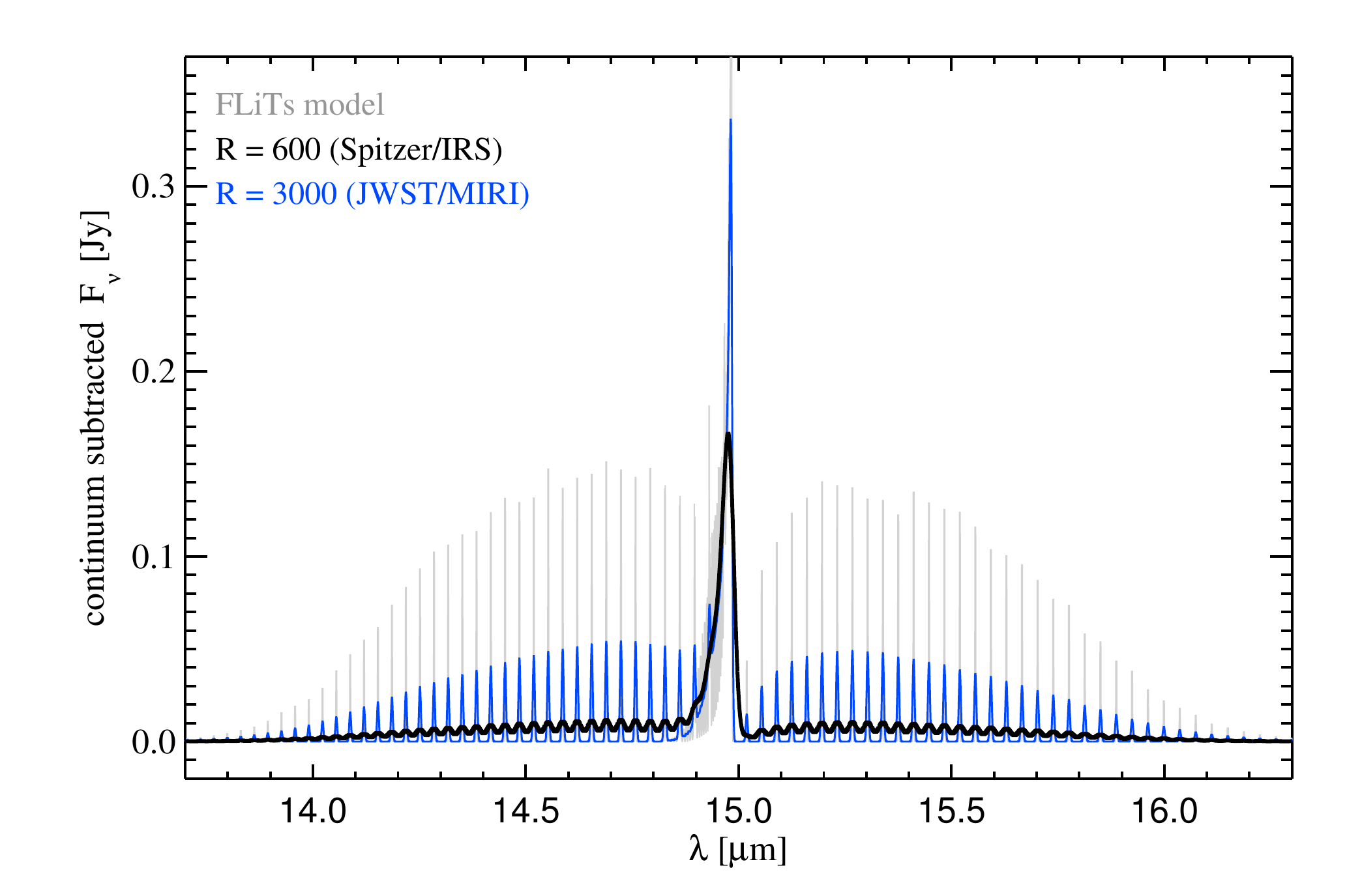}
&
\hspace*{-8mm}
\includegraphics[width=9.7cm,trim=10 11 4 4, clip]{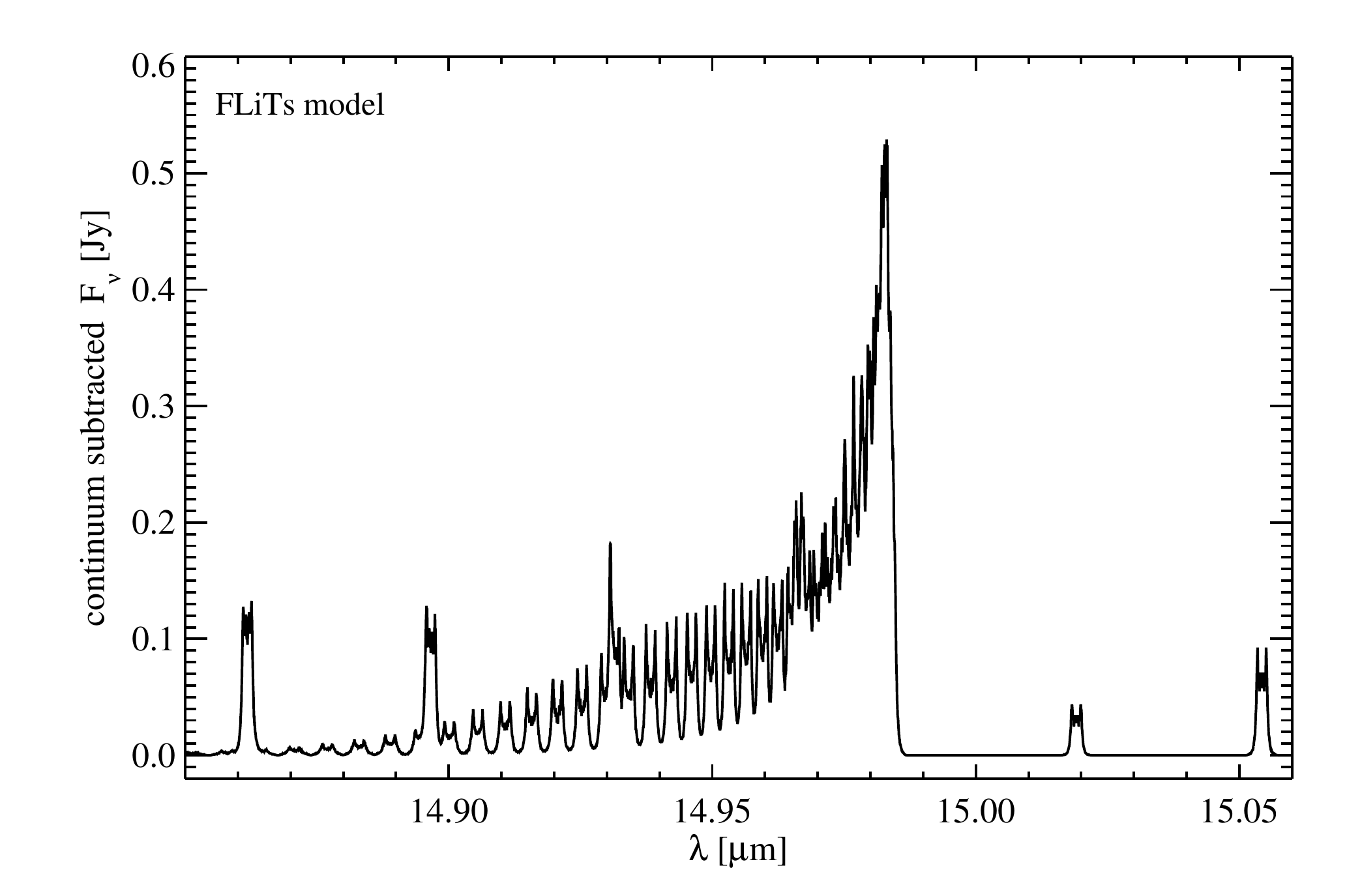}\\[-2mm]
\end{tabular}
\caption{Pure CO$_2$ $\rm 01^10(1) \to 00^00(1)$ emission spectrum
  around 15\,$\mu$m, showing the central $Q$-branch and the $P$ and
  $R$ side branches.  On the left side, the full wavelength range is
  depicted; the grey line shows the original {\sc FLiTs} spectrum, the
  black line shows the results convolved with a $R\!=\!600$ Gaussian,
  and the blue line the results convolved with a $R\!=\!3000$
  Gaussian. The right plot shows a zoom into the $Q$-branch band-head,
  plotting only the original {\sc FLiTs} spectrum.  The individual
  lines have Keplerian double-peaked profiles, and merge with each
  other at maximum. At the band head, the total flux is smaller than
  expected from the sum of the individual $Q$-branch line fluxes
  because the lines physically overlap and partly shield each other in
  the disc.}
\label{FLiTs-ProDiMo3}
\vspace*{-1mm}
\end{figure*}

The basic equations and computational techniques used for line
tracing are described in \citet[][see Appendix A7]{Woitke2011} and
\citet{Pontoppidan2009}.  Pontoppidan et al.\ developed a similar
line tracer called {\sc RadLite}. Although we used some of the
techniques and tricks described in their paper, we decided that our
specific needs for speed and flexibility require a dedicated module
for two reasons. First, we want it to be as fast and lightweight as
possible. {\sc RadLite} takes about one hour to compute 1000 lines
(3.6\,s per line); this is too long when fitting observations or
making large grids of models. Second, {\sc RadLite} traces on a
line-by-line basis. This implies that we would not be able to compute
line-blends self-consistently; see examples in
Figs.~\ref{FLiTs-ProDiMo2} and \ref{FLiTs-ProDiMo3}. For this purpose,
a new module was built from scratch that fulfils these requirements.

\subsection{Numerical implementation}

All physical quantities (i.e.\ dust opacities, dust source function with
isotropic scattering, dust and gas temperatures, molecular
concentrations, and non-LTE or LTE level populations) are passed from {\sc
  ProDiMo} to {\sc FLiTs} on the 2D spatial grid points used by {\sc
  ProDiMo}.  We assume Keplerian gas velocities, neglecting the effect
of radial pressure gradients in the disc that typically slows down
the rotation of the gas by a few percent in a power-law disc
\citep[e.g.][]{Tazaki2015}. \citet{Pinte2018} have recently provided
strong observational evidence that the disc of IM Lupi rotates with
sup-Keplerian velocities beyond the tapering-off radius, where the
column density decreases exponentially, but not at radial distances
relevant to the IR molecular emission lines.  In order to use the
computational accelerations described in \citep{Pontoppidan2009}, we
have to transform the point-based {\sc ProDiMo} grid into a cell-based
grid. Although we try to avoid interpolations as much as possible,
this transformation can introduce some minor differences between the
lines directly computed by {\sc ProDiMo} and the lines computed by
{\sc FLiTs}.

To calculate the line spectra, we integrate the monochromatic formal
solution of radiative transfer for continuum\,$+$\,lines along
multiple parallel rays through the disc, at given inclination angle,
for each wavelength. This way we construct an image of the disc at
each wavelength (see Fig.~\ref{FLiTs-ProDiMo1}). One of the crucial
parts in this procedure is to avoid aliasing effects from the way the
spatial or spectral grid are sampled. As we detail below, we avoid
this aliasing by efficiently randomising the spatial and spectral
sampling. This is very similar to using Monte Carlo integration
techniques to integrate over the spatial extent of the disc and the
finite width of a spectral bin.

In principle we need to carefully sample the physical line
profile function given by thermal $+$ turbulent broadening, using a high
enough spectral resolution. However, we created {\sc FLiTs} in such a
way that the model is also able to produce accurate results when running in
significantly reduced spectral resolution. To do this, we do not
sample each ray at exactly the same wavelength, but for each ray
tracing through the disc we take a slightly different wavelength,
randomly chosen within the wavelength bin considered. This way we randomly 
integrate over the finite width of the wavelength bin.

To increase the speed further, we store all computations done to avoid
repeating the same work twice. This applies for example to the line
profile function for each molecule at each location in the disc, and
the continuum source function at each location and wavelength. At each
velocity bin, the line contribution originates only from a very small
region of the disc image. This means that we can reduce the
computation time by finding exactly where that region is, and use the
solution for the continuum ray tracing for the remaining disc
image. For each parallel ray we store which cells are passed and what
the projected velocity is. This way we know which rays contribute to
which part of the observed line profile.

\begin{table}[!b]
\vspace*{0mm}
\caption{Integrated line fluxes computed by {\sc FLiTs} for the
  o-H$_2$O rotational line $\rm 13_{\,6,5}\to 12_{\,4,9}$ at
  12.453\,$\mu$m\ \ [$10^{-18}$\,W/m$^2$]$^{(\star)}$\ \ as a function of
  velocity channel width $\Delta v$ and accuracy settings, based on
  the $140\times140$ disc model.}
\label{tab:linefluxes}
\vspace*{-3.5mm}
\begin{center}
\def\z{\hspace*{-1mm}}
\vspace*{-2mm}
\begin{tabular}{c|cccc}
\hline
\z Accuracy setting   & $-1$ (lowest)\z & \z$1$ (default)\z 
                   & \z$5$ (highest)\z \\
\hline
&&&\\[-2ex]
$\Delta v=10$\,km/s & $4.56 \pm 0.48$  & $4.80 \pm 0.15$ & $4.88 \pm 0.07$ \\
$\Delta v=5$\,km/s  & $4.81 \pm 0.25$  & $4.84 \pm 0.06$ & $4.86 \pm 0.04$ \\
$\Delta v=1$\,km/s  & $4.83 \pm 0.04$  & $4.86 \pm 0.02$ & 
                                           \z$4.86 \pm 0.003$\z\z\\ 
\hline
\end{tabular}
\end{center}
{\ }\\*[-3.5mm] {\footnotesize $^{(\star)}$: The numbers after $\pm$
  denote the standard deviation of the line flux as estimated from a
  number of runs with different seeds for the random number generator.
  The results are consistent with ProDiMo
  ($4.84\times10^{-18}$\,W/m$^2$); see Fig.~\ref{FLiTs-ProDiMo1}.}
\end{table}

\subsection{Spatial sampling of the rays and accuracy}

The most crucial part of the line radiative transfer is the spatial
selection of ray positions and the underlying spatial resolution of
the disc model. Each part of the disc is responsible for a different
velocity component of the lines. We need to accurately sample the disc
to resolve the line emitting regions at each wavelength, yet we have
to make sure we do not oversample to avoid unnecessary
computations. In {\sc FLiTs} the disc is sampled by a bundle of
parallel rays, the positions of which are determined by the projection
of a number of random locations within each 2D disc model cell onto
the image plane.  Since the 2D disc model grid was set up to properly
trace the temperature, density, and chemical variations in the disc,
we now automatically sample this properly as well.  The number of rays
is determined by the accuracy settings in {\sc FLiTs}. A larger
accuracy number mean that more points per 2D disc model cell are
projected onto the image plane, resulting in a better spatial
resolution of the disc. Next we create a Delaunay triangulation from
those selected points and use the centre of all triangles as ray
positions. Since the positions of the rays are randomised this way, we
avoid aliasing effects that would be visible when using relatively low
resolution fixed spatial sampling. We also create a number of sets of
ray positions and pick one of these for each wavelength to reduce the
spatial sampling problems further.

\begin{table}[!b]
\vspace*{-2mm}
\caption{Time consumption of {\sc FLiTs} in CPU sec\,/\,spectral
  line$^{(\star)}$ as a function of disc model grid size $N_r \times
  N_z$ and velocity channel width $\Delta v$.}
\label{tab:CPUtime}
\begin{center}
\def\z{\hspace*{-1mm}}
\vspace*{-5mm}
\begin{tabular}{c|ccccc}
\hline
model grid size   & $50\times 40$\z & \z$90\times 70$\z 
                  & \z$140\times 140$\z & \z$200\times 300$\z \\
\hline
&&&&\\[-2.2ex]
$\Delta v=5$\,km/s   & 0.028 & 0.061 & 0.18 & 0.71 \\
$\Delta v=2$\,km/s   & 0.066 & 0.14  & 0.40 & 1.5 \\
$\Delta v=1$\,km/s   & 0.13  & 0.28  & 0.80 & 3.0 \\
$\Delta v=0.5$\,km/s & 0.29  & 0.58  & 1.6  & 5.9 \\
\hline
\end{tabular}
\end{center}
{\ }\\*[-3.5mm] {\footnotesize $^{(\star)}$: Tested on a 3.4\,GHz
  Linux computer with accuracy$=\!-1$. The `accuracy' setting in
    {\sc FLiTs} determines how many rays are used to sample the disc,
    depending on the original {\sc ProDiMo} disc model grid size.
  For the lowest accuracy level chosen here, {\sc FLiTs} uses about
  3000 rays for the $50\times 40$ model, 5000 rays for $90\times 70$,
  7000 rays for $140\times 140$, and 12000 rays for $200\times
  300$. For higher accuracy settings, more rays are used to
  better resolve the disc in the image plane.}
\end{table}

In Fig.~\ref{FLiTs-ProDiMo1} we show a comparison of the computed
single line profiles, demonstrating that the line profiles computed
by {\sc ProDiMo} are reproduced by {\sc FLiTs}, with no systematic
differences due to interpolation or numerics.
Table~\ref{tab:linefluxes} shows the computed line fluxes for one
selected water line for various settings of spectral resolution and
accuracy.  The line fluxes computed by {\sc FLiTs} show no systematic
errors even for the lowest accuracy and poorest spectral
resolution. However, higher accuracies are required to obtain precise
line profiles as shown in Figs.~\ref{FLiTs-ProDiMo1},
\ref{FLiTs-ProDiMo2}, and \ref{FLiTs-ProDiMo3} (right side) to
eliminate the noise introduced by the random spectral and spatial
sampling.

Figure \ref{FLiTs-ProDiMo3} represents how the superior
signal-to-noise ratio and the improved spectral resolution of
JWST/MIRI allows us for the first time to resolve individual $P$-
and $Q$-branch lines of molecules like CO$_2$ in the mid-IR, and to 
use the ratios of those line fluxes as a thermometer, as already 
proposed by \citet{Bosman2017}.

\subsection{Computational time requirements}

The computational time consumption of {\sc FLiTs} depends on the
accuracy and velocity resolution requested. Since the code is mainly
developed for medium excitation ro-vibrational molecular lines with
application to low spectral resolution observational mid-IR data, we
do not need high accuracy nor velocity resolution for the subsequent
models presented in this paper. Table~\ref{tab:CPUtime} lists the
computational time requirements of {\sc FLiTs} when simulating a large
number of lines with low accuracy setting. Noteworthy, the line fluxes
obtained depend on the {\sc ProDiMo} disc model grid size \citep[see
  Appendix~E in][]{Woitke2016}, which is related to the difficulty to
spatially resolve the thin line forming regions; see
Sect.~\ref{sec:details}. All results presented in Sect.~6 and beyond
have been produced using $140\times 140$ disc models, {\sc FLiTs}
accuracy$\,=\!-1$ and $\Delta v\!=\!5\,$km/s, which corresponds to
about 0.18 CPU-sec/line and a line flux error of about 5\%. In
contrast, the high spectral resolution results shown in
Figs.\ \ref{FLiTs-ProDiMo1} to \ref{FLiTs-ProDiMo3} have been produced
with accuracy\,$=\!+1$ and $\Delta v\!=\!1\,$km/s.

\section{Selection of molecules, levels, and lines}
\label{sec:Levels}

\begin{table}
\vspace*{1mm}
\caption{Selected atoms and molecules in the mid-IR spectral
  region.}
\label{tab:IRdata}
\def\z{\hspace*{-1mm}}
\def\zz{\hspace*{-4mm}}
\vspace*{-3mm}
\hspace*{-1mm}\resizebox{9.1cm}{!}{
\begin{tabular}{l|c|ccc|c}
\hline
&&&&&\\[-2.2ex]
        & Treatment    & $\lambda\rm\,[\mu m]$ & \#Levels$^{(9)}$ & \#Lines 
        & Reference$\!\!\!\!$\\
\hline
$\rm H_2O$      & non-LTE & $2.3-600$ & 824  & 8190 & (1),(2),(4)\\
$\rm OH$ rot.   & non-LTE & $25-120$  &  20  & 50   & (3),(8)\\
$\rm OH$ ro-vib.\z& HITRAN& $10-50$   & 2528 & 1264 & (4)\\
$\rm CO_2$      & HITRAN  & $13-17$   &  252 & 126  & (4)\\
$\rm HCN$       & HITRAN  & $12-17$   &  252 & 126  & (4)\\  
$\rm C_2H_2$    & HITRAN  & $11-16$   & 1992 & 996  & (4)\\
$\rm NH_3$ ro-vib.\z&HITRAN& $9-50$   & 5932 & 2966 & (4)\\
$\rm CH_4$      & HITRAN  & $18-25$   &  430 & 215  & (4)\\
$\rm NO$        & HITRAN  & $28-50$   &  372 & 186  & (4)\\
$\rm H_2CO$     & HITRAN  & $19-50$   & 3134 & 1567 & (4)\\
$\rm CH_3OH$    & HITRAN  & $9-11$    &28570 &14285 & (4)\\
$\rm SO_2$      & HITRAN  & $7-10$    &41590 &20795 & (4)\\
$\rm H_2S$      & HITRAN  & $6-10$    & 1102 &  551 & (4)\\
$\rm H_2$       & non-LTE & $0.3-29$  &  160 & 1539 & (6),(7)\\
$\rm Ne^+$      & non-LTE & $0.4-12.81$ &  3 &    3 & (5)\\ 
$\rm Ne^{++}$   & non-LTE & $0.18-15.55$ & 5 &    9 & (3)\\
$\rm Ar^+$      & non-LTE & $6.985$     &  2 &    1 & (3)\\ 
$\rm Ar^{++}$   & non-LTE & $0.3-8.985$ &  5 &    9 & (3)\\ 
$\rm Fe^+$      & non-LTE & $0.2-25.99$ &120 &  956 & (5)\\ 
$\rm Si^+$      & non-LTE & $0.1-34.81$ & 15 &   35 & (5)\\ 
$\rm S$         & non-LTE & $1.7-25.25$ &  3 &    3 & (3)\\ 
$\rm S^{+}$     & non-LTE & $0.4-31.45$ &  5 &    9 & (5)\\ 
$\rm S^{++}$    & non-LTE & $0.3-33.46$ &  5 &    9 & (3)\\ 
\hline
\end{tabular}}\\[1mm]
{\footnotesize
(1) \citet{Faure2008}; 
(2) \citet{Daniel2011}; 
(3) LAMDA database \citep{Lambda2005};
(4) HITRAN 2009 database \citep{Rothman1998,Rothman2013};
(5) CHIANTI database \citep{Chianti1997};
(6) \citet{Wrathmall2007};
(7) \citet{Lique2015};
(8) \cite{Offer1994};
(9) for HITRAN molecules, the number of levels is by construction
  equal to $2\times$ the number of lines. Usually, all available 
  levels and lines are included from the various databases, 
  within the listed wavelength intervals. However, there are
  a few exceptions as follows:\\ 
OH ro-vib: levels with $E_u=900-30000\,$K only;\\
CO$_2$: only band $01101 \to 00001$ and $E_u<5000\,$K;\\
HCN: only band $0110 \to 0000$ and $E_u<5000\,$K;\\
NH$_3$: only band $0100 \to 0000$ with $E_u=550-10000\,$K;\\
C$_2$H$_2$, CH$_4$, SO$_2$, H$_2$S: levels with $E_u<5000\,$K only;\\
  NO, H$_2$CO, CH$_3$OH: levels with $E_u<10000\,$K only.
}
\vspace*{0mm}
\end{table}

The selection of mid-IR active molecules and line lists for this paper
is described in Table~\ref{tab:IRdata}. We concentrate on molecules
that have relevant line transitions between 9\,$\mu$m and
40\,$\mu$m. The extra heating/cooling rates caused by the
absorption/emission of line photons by these species are automatically
taken into account in {\sc ProDiMo}. Other atomic and molecular
species important for the radiative heating/cooling, for example those
with fine-structure or pure rotational lines in the far-IR and at
millimetre wavelengths, are included as well
\citep[see][]{Woitke2009,Kamp2010,Woitke2011}. But these other species
are not explicitly listed in Table~\ref{tab:IRdata}.

The new molecules include OH ro-vibrational, CO$_2$, HCN, C$_2$H$_2$,
NH$_3$, CH$_4$, NO, H$_2$CO, CH$_3$OH, SO$_2$, and H$_2$S, for which we
take the level energies, degeneracies, Einstein coefficients, line
centre frequencies, and partition functions from the HITRAN
database \citep{Rothman1998,Rothman2013}.  In this paper, we assume
that the levels of these HITRAN molecules are populated in LTE,
i.e. given by a Boltzmann distribution
\begin{equation}
  n_k^{\rm LTE} = n_{\rm sp} \frac{g_k}{Q(T_{\rm gas})}
  \exp\bigg(-\frac{E_k}{kT_{\rm gas}}\bigg) \ ,
\end{equation}
where $n_{\rm sp}$ is the total molecular particle density taken from
the chemistry, $n_k\,\rm[1/cm^3]$ the level population, $g_k$ the level
degeneracy, $E_k$ the level energy, and $Q(T)$ the partition
function. $T_{\rm gas}$ is the gas temperature calculated in
heating/cooling balance.

\begin{figure*}
\centering
\vspace*{-4mm}
\hspace*{-3mm}
\includegraphics[width=17cm,height=12.1cm]{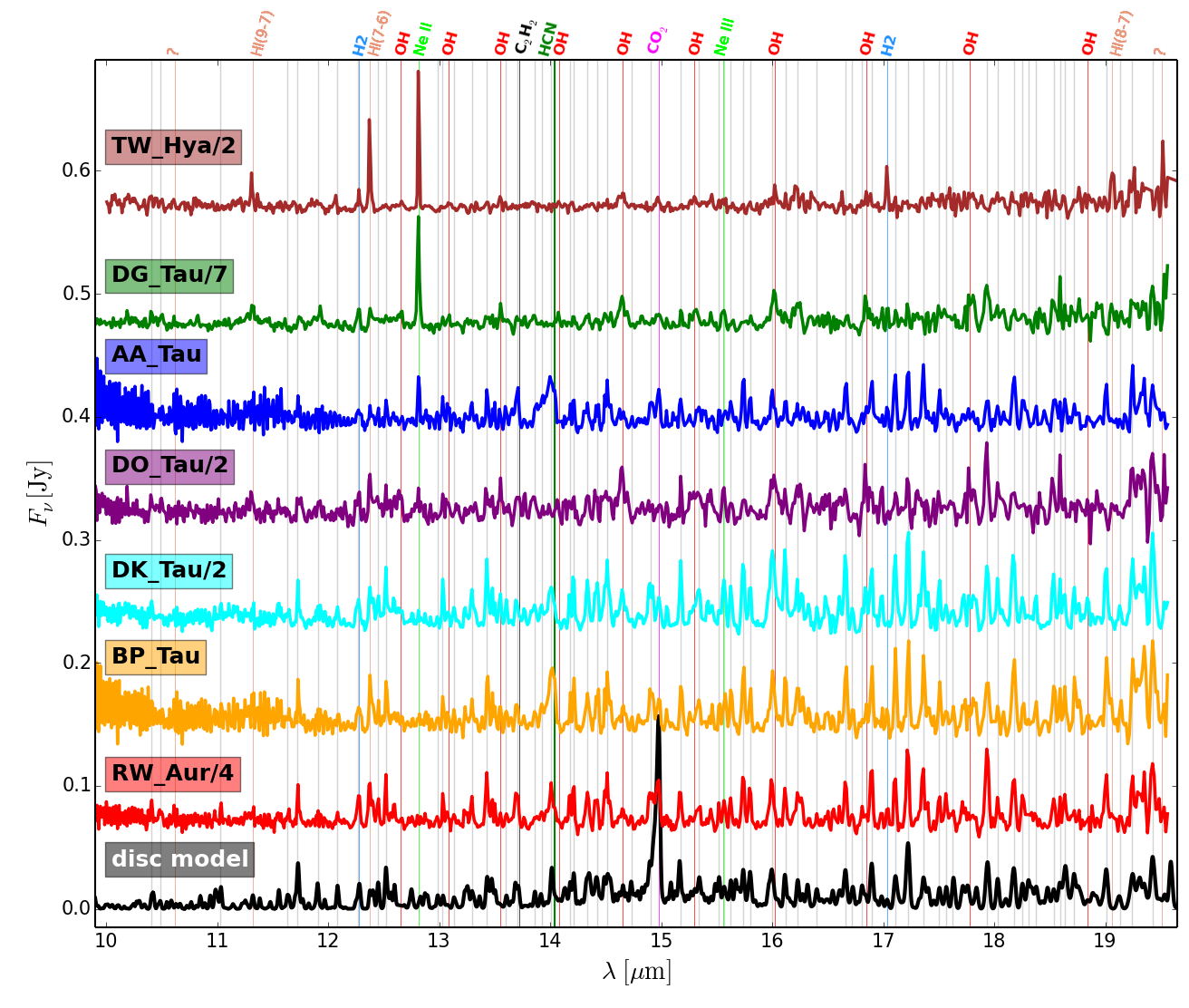}\\[-7mm]
\hspace*{-3mm}
\includegraphics[width=16.95cm,height=6.0cm]{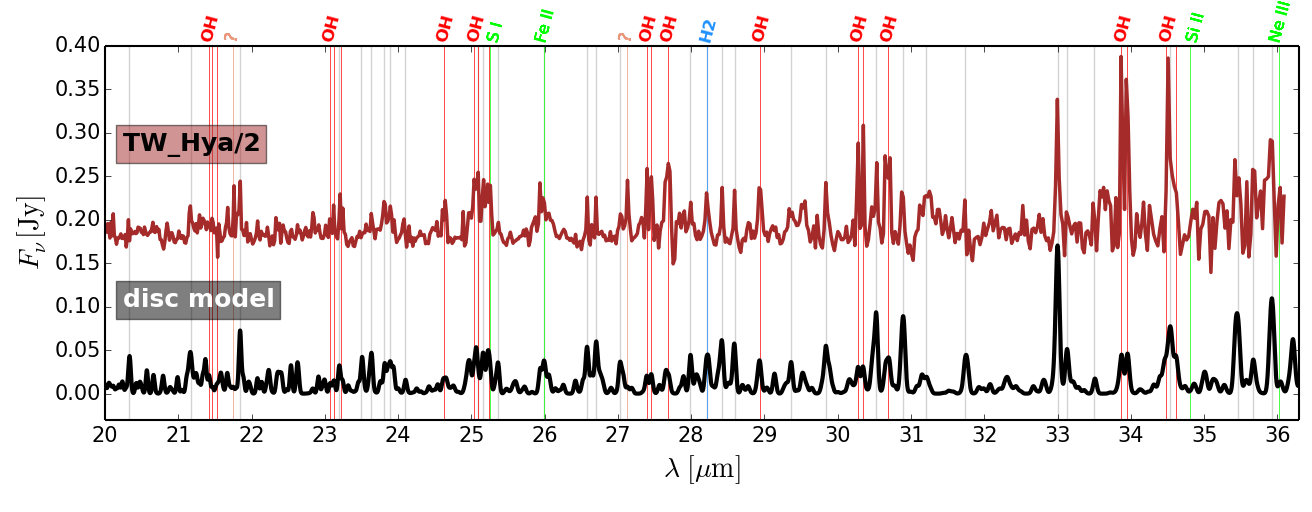}\\[-4.5mm]
\caption{{\bf Upper plot:} Seven continuum-subtracted $R\!=\!600$
  Spitzer/IRS spectra of T\,Tauri stars (coloured lines) from Zhang et
  al.\,(2013, TW\,Hya) and Rigliaco et al.\,(2015, all other objects),
  arbitrarily shifted and scaled as indicated. At the bottom of the
  upper plot, the continuum-subtracted {\sc ProDiMo\,/\,FLiTs}
  spectrum of our main disc model with $\rm gas/dust=1000$ is shown in
  black (convolved to $R=600$). {\bf Lower plot:} The model spectrum
  is continued for longer wavelengths and compared to the observations
  of TW\,Hya with strong OH lines. The thin vertical coloured lines
  and top labels identify the molecules and ions. All unlabelled grey
  vertical lines indicate water lines. The salmon-coloured lines have
  no counterpart in the model; they are either high-excitation neutral
  hydrogen lines as indicated or are unidentified when labelled with
  ``?''.}
\label{fig:commonlines}
\vspace*{-1mm}
\end{figure*}

As we can handle only a few tens of thousands of spectral lines with
{\sc ProDiMo} and {\sc FLiTs}, we need to select the
relevant parts of the line lists in HITRAN, which originally contain
many millions of transitions for a number of isotopologues.  Our
selection criteria are currently adjusted to the detections of the
Spitzer Space Telescope, see Fig.~\ref{fig:commonlines}, but this
could easily be changed.

Our current selection of lines is restricted to particular vibrational
bands and wavelength intervals, and only lines from the main
isotopologues are taken into account; see details in 
Table~\ref{tab:IRdata}. The following additional selection
rule about the strength of the lines is applied to all HITRAN
molecules to further limit the computational efforts in our disc models
\begin{equation}
  A_{ul} g_u \exp\left(-\frac{E_u}{k\cdot 1500\,{\rm K}}\right) 
  \ > \ 10^{-5}{\rm s}^{-1} \quad ,
  \label{eq:strength}
\end{equation}
where $A_{ul}$ is the Einstein coefficient, $E_u$ is the upper level
energy, and $g_u$ is the upper level degeneracy. Using
Eq.~(\ref{eq:strength}) for the line selection was an important step
to construct feasible models that are sufficient to predict all
observed emission features detected by {\sc Spitzer/IRS}.  

\begin{figure*}
\centering
\vspace*{-3mm}
\includegraphics[width=16.8cm,height=10.1cm]{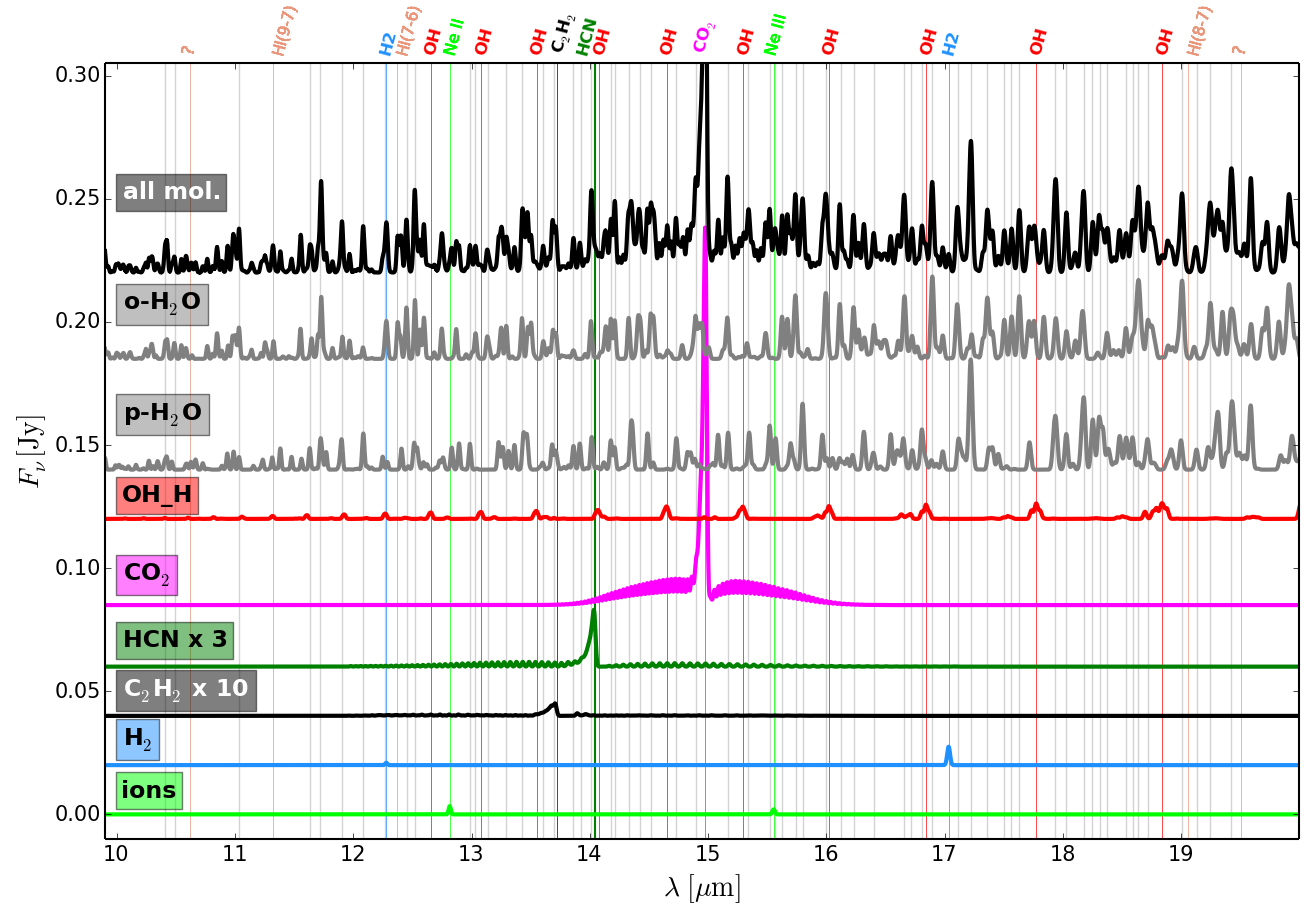}\\[-6mm]
\includegraphics[width=16.8cm,height=8.2cm]{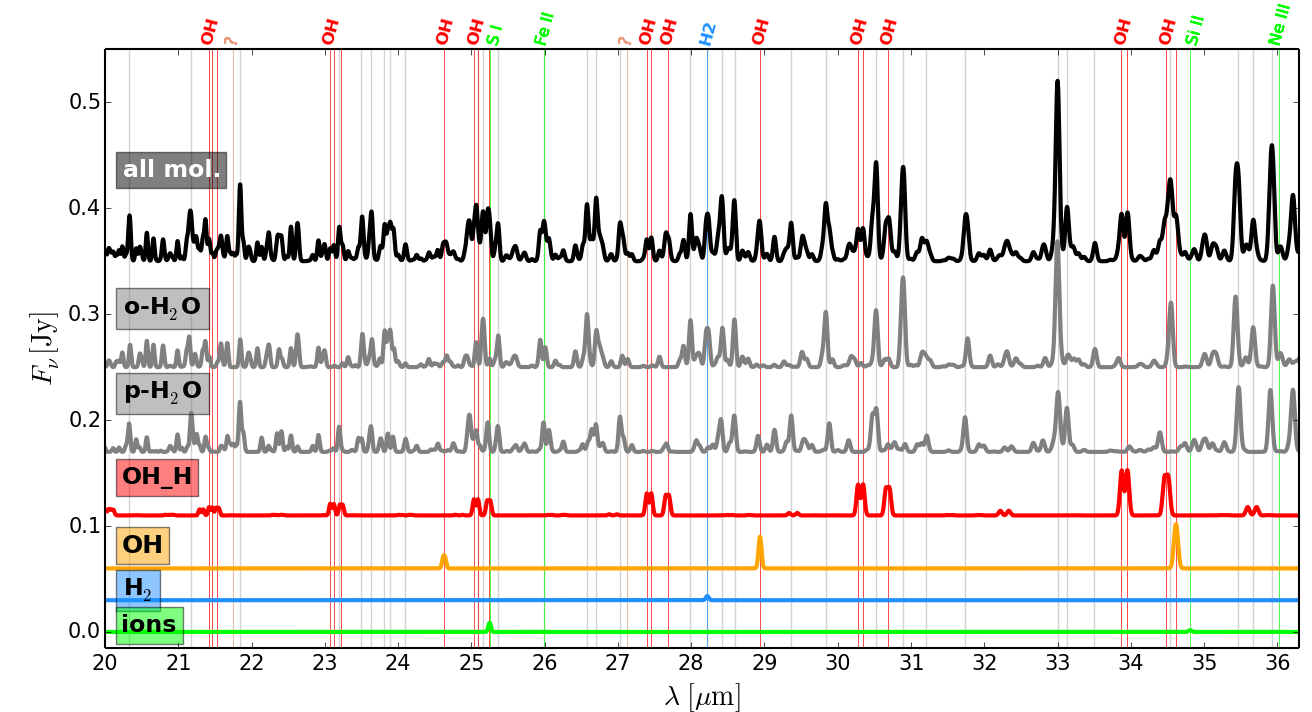}\\[-1.5mm]
\caption{The $\rm g/d\!=\!1000$ model spectrum is decomposed into its
  main molecular constituents. We use the notation ``OH\_H'' for
  ro-vibrational OH lines from the HITRAN database in contrast to the
  pure rotational lines computed in non-LTE and denoted by 
  ``OH''. These single molecule spectra are convolved to $R\!=\!600$
  and arbitrarily shifted, but not scaled except for C$_2$H$_2$ and HCN. The
  C$_2$H$_2$ lines around 13.7\,$\mu$m are very weak in the model, and
  are amplified by a factor of ten in this figure to make them visible. The
  vertical coloured lines and top labels identify the molecules and
  ions in the same way as in Fig.~\ref{fig:commonlines}.}
\vspace*{-2mm}
\label{decompose}
\end{figure*}

%============================================================================
\section{Results}
%============================================================================
\subsection{Comparison to Spitzer observations}

In Fig.~\ref{fig:commonlines}, the {\sc FLiTs} spectrum obtained from
our standard {\sc ProDiMo} disc model with gas/dust$\,=\,$1000 is
compared to the continuum-subtracted {\sc Spitzer/IRS} $R\!=\!600$
spectra of seven well-known T\,Tauri stars from \citet{Rigliaco2015}
and \citet{Zhang2013}. The figure shows numerous common emission
features in model and observations. These emission features are often
composed of several (up to hundreds of) overlapping individual lines,
in particular at shorter wavelengths. Over 100 of such common spectral
emission features (mostly water) have been identified and represented
in Fig.~\ref{fig:commonlines} by coloured vertical thin lines, where
the observational peaks have a corresponding counterpart in the model
and vice versa. We were unable to find a corresponding match with the
model for six observed lines/features. Three of these features are
high excitation neutral hydrogen atomic lines as indicated in the
figure, i.e.\ HI\,(9-7)\,11.32\,$\mu$m, HI\,(7-6)\,12.37\,$\mu$m,
HI\,(8-7)\,19.06\,$\mu$m \citep{Rigliaco2015}, which are not included
in our disc model, and three features remain unidentified at
10.62\,$\mu$m, 21.75\,$\mu$m, and 27.13\,$\mu$m. We are not claiming,
however, that our emission feature identifications are entirely
accurate. Our molecular line data are possibly incomplete, and many of
the spectral lines overlap at $R\!=\!600$ resolution; see
\citet{Pontoppidan2010} for more details.

The peak amplitudes are also in reasonable agreement with the
observations, simultaneously for different molecules. Only our
CO$_2$ emission feature at 15\,$\mu$m is too strong by about a factor
of 3. The different T\,Tauri stars show different overall levels and 
colours of line emission, and different emission feature fluxes for
different molecules. With the exception of CO$_2$ 15\,$\mu$m, the
observed scatter in those feature fluxes is larger than the
systematic deviations from the model.

\subsection{Decomposed model spectrum}

In Fig.~\ref{decompose} the model spectrum is decomposed
into its molecular constituents, confirming the following results:
\begin{itemize}
\setlength{\itemsep}{1pt}
\setlength{\parskip}{0pt}
\setlength{\topsep}{0pt}
\setlength{\parsep}{0pt}
\setlength{\partopsep}{0pt}
\item Water is the main contributor to the mid-IR line spectra
  of T\,Tauri stars \citep{Pontoppidan2010}.
\item OH lines can be equally strong at longer wavelengths, clearly
  visible in the TW\,Hya spectrum \citep{Najita2010,Zhang2013}, and
  this is true for both the rotational OH and ro-vibrational OH\_H
  lines.
\item The CO$_2$ Q-branch of the $01^10(1)\!\to\!00^00(1)$ band at about
  15\,$\mu$m is regularly detected in T\,Tauri stars
  \citep{Carr2008}. The main model overpredicts the
  strength of this feature by about a factor of 3.
\item The HCN Q-branch of the $01^10\,\to\,00^00$ band at 
  14\,$\mu$m is detected in some T\,Tauri stars
  \citep{Pascucci2009,Salyk2011,Najita2013}, but not in all
  cases. Figure~\ref{decompose} shows that this feature is blended
  with one o-H$_2$O ($12_{7,6}\!\to\!11_{4,7}$) and one p-H$_2$O
  ($10_{7,3}\!\to\!9_{4,6}$) line. In our model the HCN Q-branch
  contribution to this feature is about 50\%, {which is possibly
    somewhat too weak in comparison to some T\,Tauri star observations.}
\item The C$_2$H$_2$ feature from the 000011u\,$\to$\,000000+g system
  at about 13.7\,$\mu$m is detected in some T\,Tauri stars
  \citep{Pascucci2009}. It is very weak in the model,
  Fig.~\ref{decompose} shows it with magnification $\times 10$.  This
  is likely to be a chemical effect as our disc model does not predict
  large concentrations of C$_2$H$_2$ in the relevant disc surface
  layers, although very abundant in deeper layers; see
  Sect.~\ref{sec:details}. 
\item Some T\,Tauri stars show strong [Ne\,II]\,12.81\,$\mu$m and
  [Ne\,III]\,15.55\,$\mu$m lines \citep{Pascucci2007,Gudel2010} and
  some H$_2$\,17.03\,$\mu$m and 28.22\,$\mu$m emission, whereas others
  do not \citep{Lahuis2007,Baldovin-Saavedra2011}. In the model, we
  need particular phyical conditions to excite these lines, such as 
  vertically extended low-density regions above the disc for the Ne
  lines. In the main model, these lines are rather weak
  ($1.3\times10^{-18}\rm\,W/m^2$ and $6.6\times10^{-19}\rm\,W/m^2$ for
  the main Ne\,II and Ne\,III lines), which agrees with many T\,Tauri
  observations \citep{Aresu2012}.
\item The Fe\,II fine-structure line at $25.99\,\mu$m is
  strongly blended with a number of water lines. The line is actually very
  weak in the model $(3\times 10^{-20}\rm\,W/m^2)$, as we are using a
  very low Fe element abundance in our model, assuming that Fe is
  locked in grains.
\end{itemize}

\subsection{Role of dust/gas mass ratio and 
            $T_{\rm gas}\!>\!T_{\rm dust}$}
\label{sec:gas_dust}

\begin{figure}
\vspace*{0mm}
\hspace*{-3mm}\includegraphics[width=95mm,trim=0 100 0 0,clip]{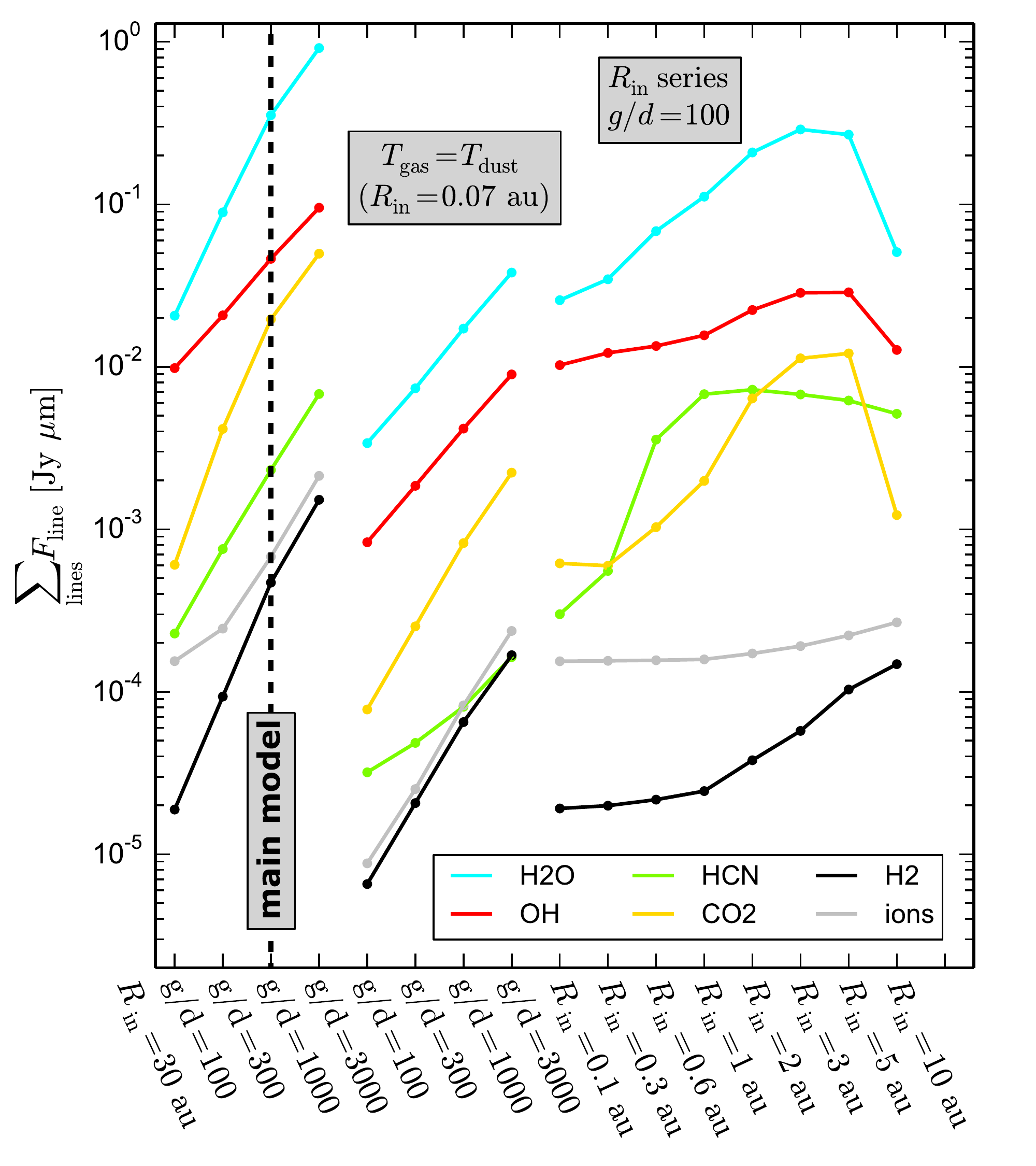}\\*[-2mm]
\hspace*{-3mm}\includegraphics[width=95mm,trim=0 0 0 8.5,clip]{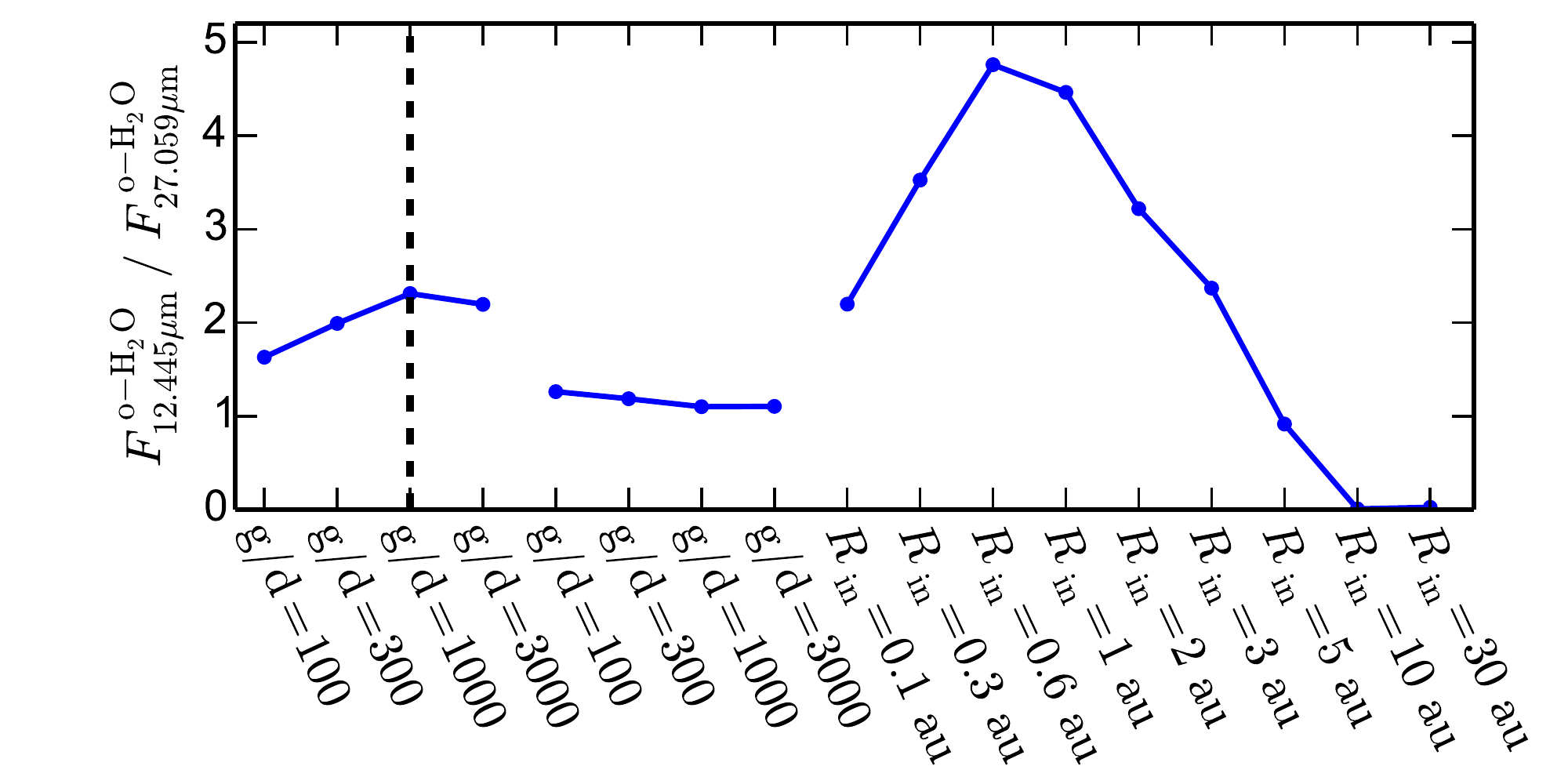}\\*[-5mm]
\caption{{\bf Upper part:} Sum of all line fluxes emitted by the
  different molecules between 9.7\,$\mu$m and 38\,$\mu$m.  {The
    vertical dashed line indicates the main model as plotted in
    Figs.~\ref{fig:commonlines} and \ref{decompose}, which is broadly
    consistent with the observations.}  {\bf Lower part:} Colour of
  the water line emission spectrum in the form of a ratio of two
  o-H$_2$O lines at about 12\,$\mu$m and 27\,$\mu$m. In the series of
  4 models on the left, the gas/dust mass ratio in the disc is varied,
  the calculated gas temperatures are used, and $R_{\rm
    in}\!=\!0.07\,$au is assumed.  In the series of 4 models in the
  centre, the gas/dust mass ratio in the disc is varied while $T_{\rm
    gas}\!=\!T_{\rm dust}$ and $R_{\rm in}\!=\!0.07\,$au are assumed.
  In the series of models on the right, the disc inner radius $R_{\rm
    in}$ is varied while assuming $\rm g/d\!=\!100$ and using the
  calculated gas temperatures. ``H2O'' is the combined emission from
  o-H$_2$O and p-H$_2$O, and ``OH'' is the combined emission from
  rotational OH lines and ro-vibrational OH\_H lines; see
  Table~\ref{tab:IRdata}.}
\label{fig:series}
\vspace*{-2mm}
\end{figure}

Figure~\ref{fig:series} shows the total fluxes of all spectral
lines emitted by the various molecules between 9.7\,$\mu$m and
38\,$\mu$m as a function of gas/dust ratio (g/d) in the model (at
constant dust mass); see series of four models on the left side in
Fig.~\ref{fig:series}. There is a very clear correlation for all
molecules. As g/d increases, larger columns of gas are present above
the $\tau_{\rm dust}\!=\!1$ surface, which leads to stronger emission
lines. An alternative explanation can be provided by studying the
heating/cooling balance. For larger g/d, more of the heating UV
photons are absorbed by the gas, rather than by the dust, and this
increase of gas heating is balanced by an increase of line cooling in
the mid-IR spectral region. The measured dependencies of line fluxes
versus g/d are about linear. A similar increase of all line fluxes can
be produced in the model by changing some of the dust size and
material properties such that the dust has lower UV and IR opacities.
 
The subsequent series of four models in Fig.~\ref{fig:series} shows the
results if we ignore the gas energy balance and artificially assume
$T_{\rm gas}\!=\!T_{\rm dust}$. In this case, the line fluxes drop by
about one order of magnitude with respect to a full model with the
same g/d, underlining the importance of the gas over dust 
  temperature contrast in the line emitting regions, as was already
shown by \citet[][]{Carmona2008} concerning H$_2$ lines and by
\citet{Meijerink2009} concerning H$_2$O lines. As we show in
Table~\ref{tab:coldens}, the overwhelming part of the observable
molecular lines are optically thick and form on
top of an optically thick dust continuum. For such lines, the
Eddington-Barbier approximation 
\begin{equation}
  F_{\rm line} \propto S_\nu^{\rm line}(\tau_{\rm line}\!=\!1) -
  B_\nu\big(T_{\rm dust}(\tau_{\rm dust}\!=\!1)\big)
\label{eq:EddBarb}
\end{equation}  
explains why the temperature contrast between gas and dust is so
important. If we consider LTE, $S_\nu^{\rm line}(\tau_{\rm
  line}\!\!=\!\!1) = B_\nu\big(T_{\rm gas}(\tau_{\rm
  line}\!\!=\!\!1)\big)$ is valid. The weak line fluxes obtained from
the $T_{\rm gas}\!=\!T_{\rm dust}$ approximation then result from the
dust temperatures in the upper line-forming disc layers being slightly
higher than the dust temperatures in the optically thick midplane. If
that dust temperature slope were reversed (as expected in active discs
powered by viscous heating), we would see absorption lines. In the
full {\sc ProDiMo} models, $T_{\rm gas}\!>\!T_{\rm dust}$ is a typical
result (see Sect.~\ref{sec:details}) although this is not true for all
layers and model parameters.  The large amplification factor of about
10 finally results from the steepness of the Planck function at mid-IR
wavelengths if the temperature drops below a few 100\,K.

\subsection{Dependence on inner disc radius}
\label{sec:Rin}

An increase of the gas/dust ratio from 100 to about 1000 provides an
easy way to obtain mid-IR spectra that are broadly consistent with
{\sc Spitzer/IRS} line observations. However, it is unclear how robust
this finding is, whether this physical interpretation is correct, or
whether there are other ways to increase the emission line fluxes to
the observed level.  While looking for alternatives, we found that
larger inner disc radii also imply higher mid-IR line fluxes in
general; see also \citet{Antonellini2016}.  On the right side of
Fig.~\ref{fig:series}, we show the results of a series of $\rm
g/d\!=\!100$ models where the inner disc radius is systematically
increased from its default value of $R_{\rm in}\!=\!0.07\,$au to
values up to 30\,au.  All mid-IR line fluxes in the models are found
to increase by factors of about 4 to 20 (depending on molecule), reach
a maximum at a few au, and then decrease
again. Figure~\ref{fig:escpro} compares the line spectrum emergent
from the main model to the line spectrum from the $\rm g/d\!=\!100$
model with $R_{\rm in}\!=\!3\,$au, showing that both options result in
about the same overall line flux level. The main difference between
these spectra is the overall colour of the line emission as shown in
the lower part of Fig.~\ref{fig:series} and discussed below. The line
emission from the wall is bluer at maximum.

A similar behaviour was noticed for CO fundamental ro-vibrational
lines and explained in \citep{Woitke2016}. As the inner disc radius
$R_{\rm in}$ is increased in the model, the line emissions from the
disc surface are more and more replaced by the direct emissions from
the visible area of the inner rim of the disc; see
Fig.~\ref{FLiTs-ProDiMo1}. That wall is illuminated well by the star
that heats the gas in the wall.  Although we can only speculate
about the physical structure of gas and dust in these walls, it seems
reasonable to assume that very high gas densities are reached soon
inside these walls; therefore the line emission takes place at
unusually high densities where non-LTE effects are not likely to be
important.  As $R_{\rm in}$ increases, the size of the visible area of
the wall increases with $R_{\rm in}^2$ and this effect is more
important at first than the decrease of the wall temperature and line
source functions in the wall. Once $R_{\rm in}$ reaches about 10\,au,
however, the wall becomes too cold and loses its ability to emit in
the mid-IR, and so the line fluxes eventually diminish quickly.

The lower part of Fig.~\ref{fig:series} shows the colour of the water
emission spectrum in the form of the line flux ratio o-H$_2$O
$\rm 13_{\,6,5}\to 12_{\,4,9}$ at 12.453\,$\mu$m divided by
o-H$_2$O $8_{5,4}\!\to\!8_{2,7}$ at $\lambda\!=\!27.059\,\mu$m (same
lines as shown in Fig.~\ref{FLiTs-ProDiMo1}). These
two lines have excitation energies of 4213\,K and 1806\,K,
respectively.  Large line flux ratios correspond to a blue colour of
the water emission line spectrum.  We see that the models using the
calculated $T_{\rm gas}$ are not only brighter but also bluer
than the models assuming $T_{\rm gas}\!=\!T_{\rm dust}$. As $R_{\rm
  in}$ is increased in the model, the colour gets bluer first, as long
as the wall emission continues to replace the disc surface emission,
but then the colour becomes redder again as the temperature in the
distant inner wall decreases. Interestingly, \citet{Banzatti2017}
performed an observational study of water lines combined with inner
disc radii obtained from high-resolution ro-vibrational CO lines, which
show that the H$_2$O lines disappear from blue to red with
increasing disc radius.

% 138 ->   84 lam[mic]= 1.245346E+01 Eu[K]= 4.213E+03 Aul[1/s]= 1.160E+00 Ncr[cm^-2]= 3.521E+14     0    0    0,    13    7    6->     0    0    0,    12    4    9
%  40 ->   28 lam[mic]= 2.705872E+01 Eu[K]= 1.806E+03 Aul[1/s]= 4.545E-02 Ncr[cm^-2]= 9.460E+14     0    0    0,     8    5    4->     0    0    0,     8    2    7

The lines of atomic ions and H$_2$ behave in different
ways than the molecular lines discussed so far. The ion lines need an
extended, tenuous, ionised, and hot layer high above the disc to become
strong, whereas the excitation mechanism of the H$_2$ quadrupole lines
is more complex \citep[see][]{Nomura2005,Nomura2007} and depends on
wavelength.  The H$_2$ lines at longer wavelengths form deep inside
the disc in our models. Therefore, strong H$_2$ lines require large
column densities and a sufficiently large temperature contrast $T_{\rm
  gas}\!>\!T_{\rm dust}$ well inside the disc, which is usually not
present in our models.

More detailed investigations are required to study the
spectroscopic differences between the line emission spectra obtained
from disc walls and from disc surfaces; see Fig.~\ref{fig:escpro}. It
seems possible that we can distinguish between disc surface and
wall emission and that we can use the spectroscopically
  unresolved mid-IR spectroscopic fingerprints of wall emission to
detect the presence of secondary irradiated disc walls with JWST at
distances of a few au, for example caused by disc-planet interactions; see
discussion in Sect.~\ref{sec:walls}.

\subsection{Chemical structure and line origin}
\label{sec:details}

\begin{figure*}
\vspace*{-1mm}
\centering
\begin{tabular}{cc}
\includegraphics[width=9cm,trim=30 27 120 85,clip]{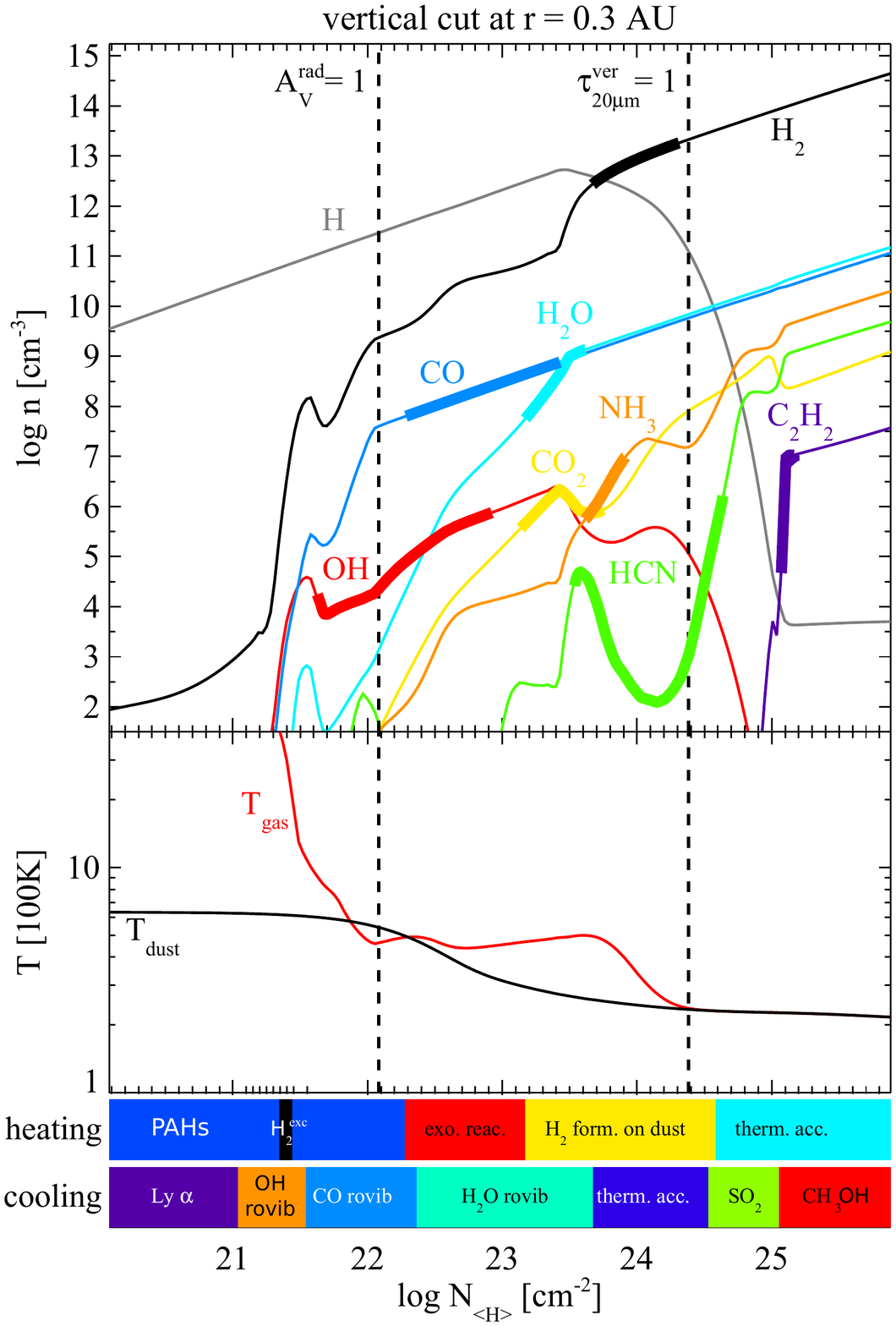}
\includegraphics[width=9cm,trim=30 27 120 85,clip]{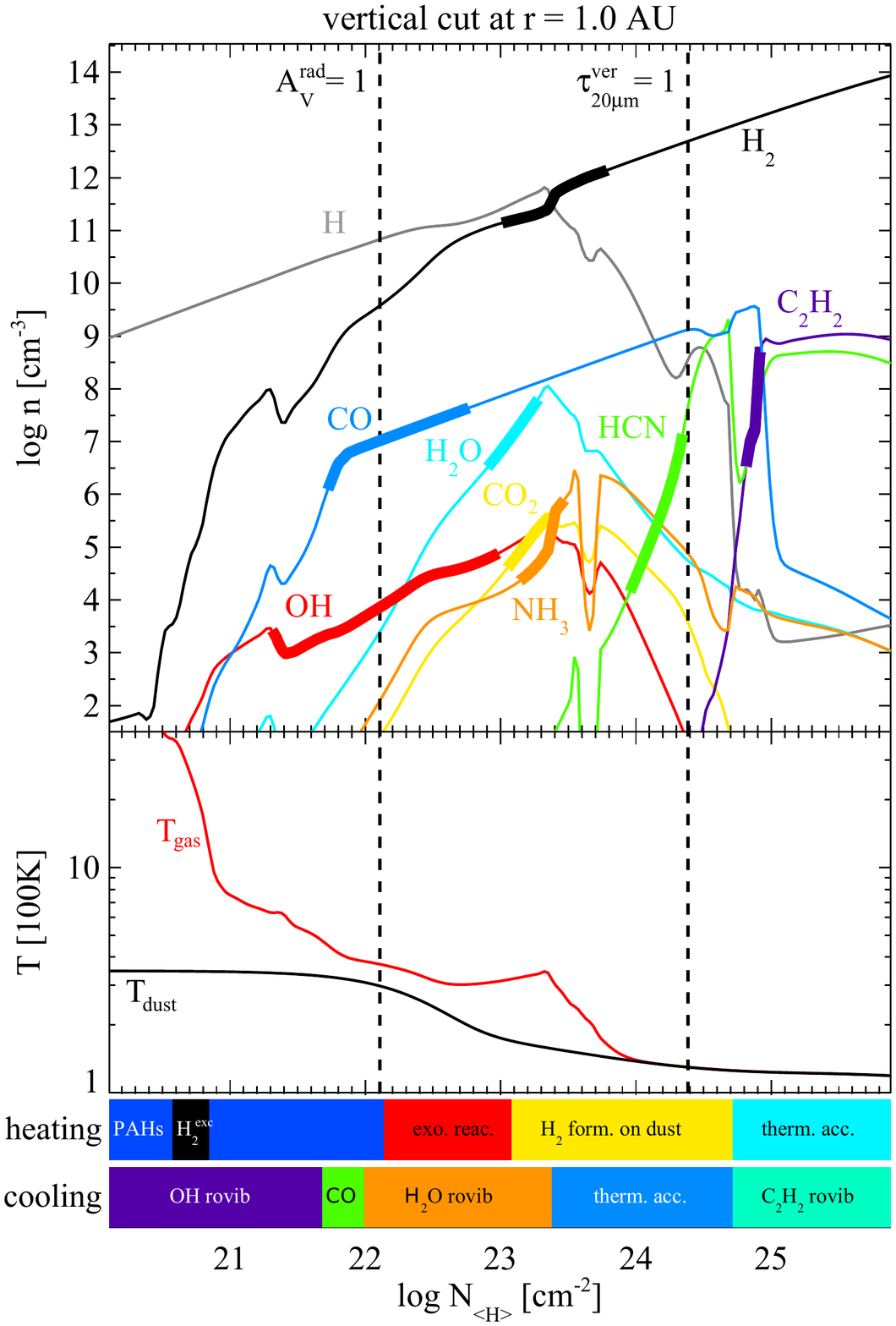}
\end{tabular}\\[-2mm]
\caption{Abundances of mid-IR active molecules along vertical cuts
  through the disc at 0.3\,au (left) and 1\,au (right), plotted as
  a function of the vertical hydrogen nuclei column density $N_\H$.  The
  thick parts of the graphs highlight the mid-IR line forming regions
  of the respective molecules (see Eq.~(\ref{eq:mean}) and text). The
  two vertical dashed lines indicate where the radial dust optical
  depths is one ($A_{\rm V}^{\rm rad}=1$), and where the vertical dust
  optical depth at $\lambda=20\,\mu$m is one. The lower plot shows the
  vertical gas and dust temperature structure in the disc. The
  coloured bars below indicate the most important heating and cooling
  processes: rovib $=$ ro-vibrational, therm.\,acc. $=$ thermal
  accommodation, exo.\,reac.~$=$ exothermal chemical reactions,
  form.~$=$ formation.}
\label{fig:vertical}
\vspace*{-2mm}
\end{figure*}

Figure~\ref{fig:vertical} shows two vertical cuts through the main
model with $\rm g/d\!=\!1000$ at $r\!=\!0.3\,$au and $r\!=\!1\,$au,
using the output from a $200\times300$ model. The vertical structure
is typical for photon dominated regions \citep[PDRs; see
  e.g.][]{Roellig2007}, where the formation of molecules is suppressed
by photodissociation at low column densities and triggered by
subsequent UV dust absorption and molecular shielding. The resulting
concentrations in discs over vertical $N_\H$ can be compared, for example to
\citep{Najita2011}. With respect to standard interstellar conditions,
we find the following main differences in the planet-forming regions
of protoplanetary discs:\\*[-2.7ex]
\begin{itemize}
\setlength{\itemsep}{1pt}
\setlength{\parskip}{0pt}
\setlength{\topsep}{0pt}
\setlength{\parsep}{0pt}
\setlength{\partopsep}{0pt}
\item[1)] much higher densities,
\item[2)] intense UV irradiation under acute angles,
\item[3)] intense IR radiation emitted from the warm dust in the disc.
\end{itemize}
In particular, (2) implies that the classical $A_V$ scale does not fully
represent the full 2D dust absorption and scattering effects for the UV
penetration, which we use in our models. We chose to depict the
vertical chemical and temperature structure as function of
the vertical hydrogen nuclei column density $N_\H$ and are
highlighting the radial $A_V^{\rm rad}\!=\!1$ layer in
Fig.~\ref{fig:vertical}, where the dust temperature starts to smoothly
decline by an overall factor of about 3. The regions above the
$A_V^{\rm rad}\!=\!1$ layer can be reached by stellar photons in a
single flight, whereas the regions below are UV-illuminated mainly by
photons scattered downward by the dust particles in the $A_V^{\rm
  rad}\!\approx\!1$ layers, with subsequent vertical dust
absorption. Despite these differences, our vertical disc cuts resemble
the following principal results of PDR modelling:
\begin{itemize}
\setlength{\itemsep}{1pt}
\setlength{\parskip}{0pt}
\setlength{\topsep}{0pt}
\setlength{\parsep}{0pt}
\setlength{\partopsep}{0pt}
\item molecular concentrations increase outside-in by many orders of
  magnitude before they typically reach some constant levels at 
  relatively model-independent column densities,
\item abundant molecules such as H$_2$ and CO form first because of
  their ability to self-shield,
\item OH needs to form before H$_2$O can form,
\item CO$_2$ forms in combination with H$_2$O,
\item HCN and NH$_3$ form below H$_2$O and CO$_2$, and
\item pure hydro-carbon molecules such as C$_2$H$_2$ are not abundant
  in the upper disc layers when the carbon to oxygen element ratio in
  the gas is $\rm C/O\!<\!1$.
\end{itemize}
The principal mechanisms that lead to such a sequence of
increasing molecular complexity with depth in discs are
described elsewhere \citep[see e.g.][]{Kamp2004,Nomura2005,Najita2011}.

\begin{figure}
\vspace*{1.5mm}
\hspace*{-6mm}\begin{tabular}{r}
\includegraphics[height=27mm,width=95mm,
                 trim=15 38 20 290,clip]{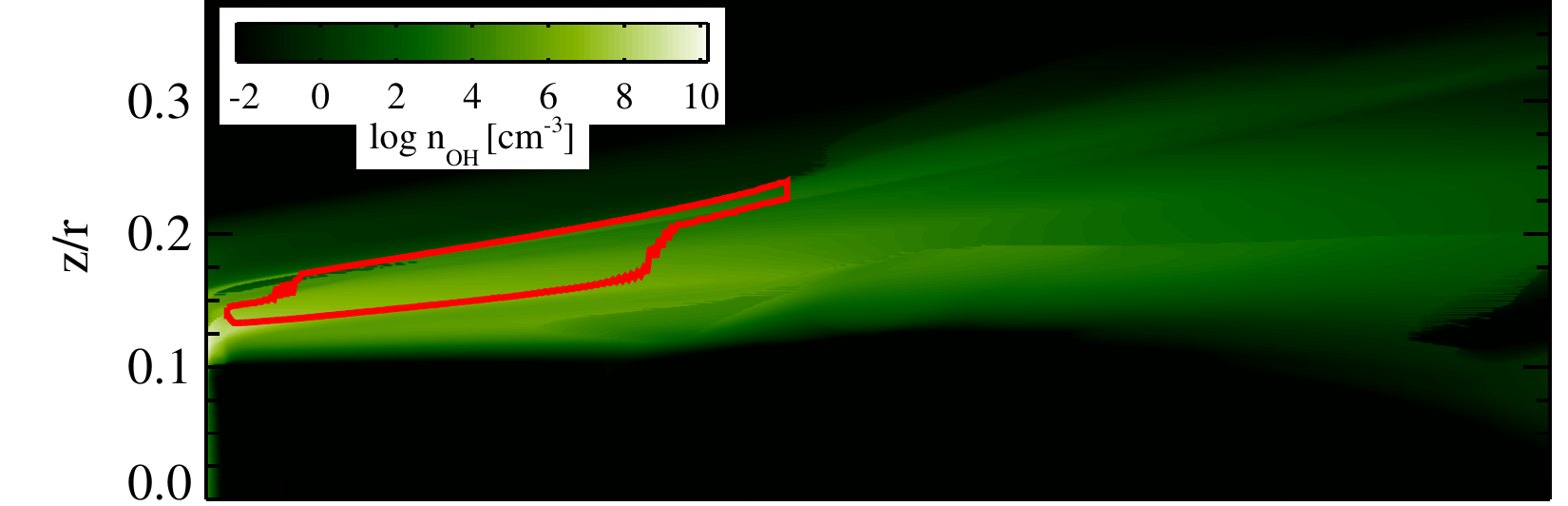}\\[-27mm]
{\color{white}\sf OH\_H $\sf \lambda\!=\!20.1151\,\mu$m}
\hspace*{15mm}\\[22.3mm]
\includegraphics[height=27mm,width=95mm,
                 trim=15 38 20 290,clip]{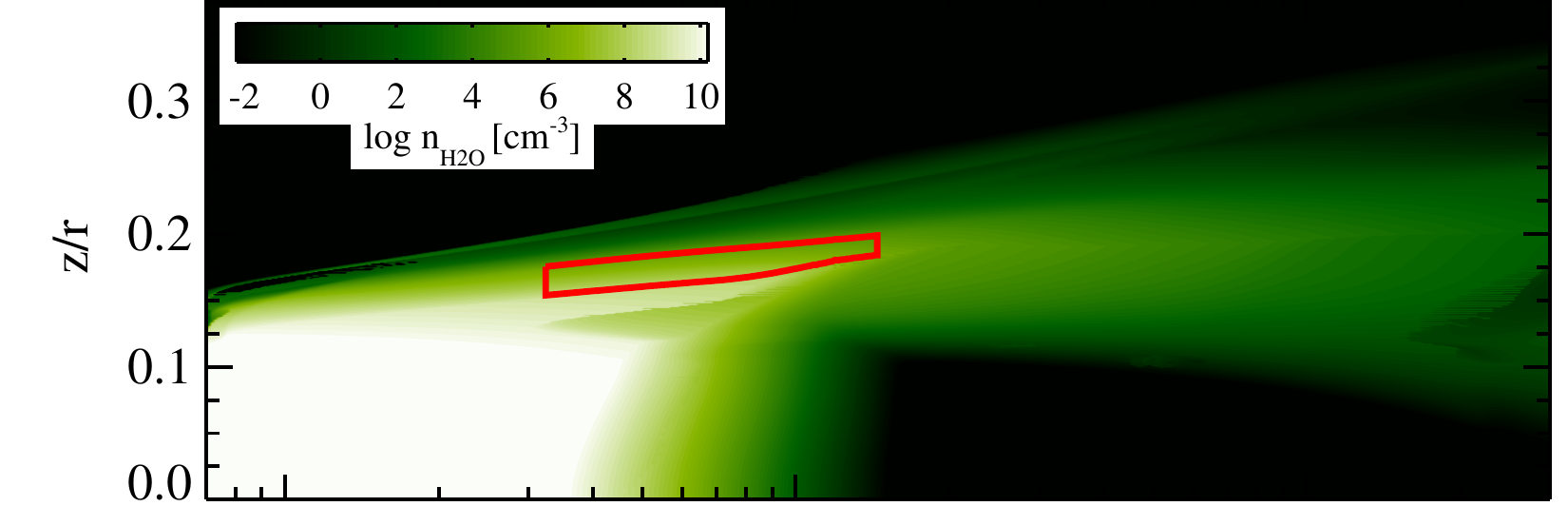}\\[-27mm]
{\color{white}\sf o-H$_{\sf 2}$O\ \ $\sf \lambda\!=\!23.8598\,\mu$m}
\hspace*{15mm}\\[22.3mm]
\includegraphics[height=27mm,width=95mm,
                 trim=15 38 20 290,clip]{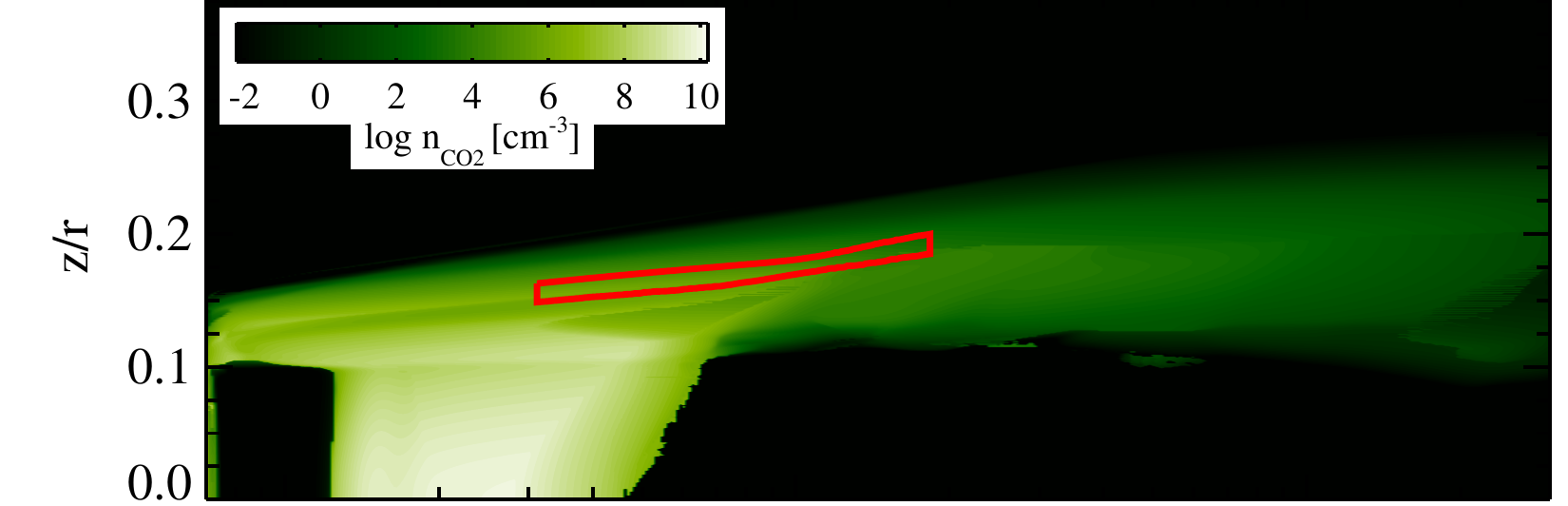}\\[-27mm]
{\color{white} CO$_{\sf 2}$ $\sf \lambda\!=\!14.9777\,\mu$m}
\hspace*{15mm}\\[22.3mm]
\includegraphics[height=27mm,width=95mm,
                 trim=15 38 20 290,clip]{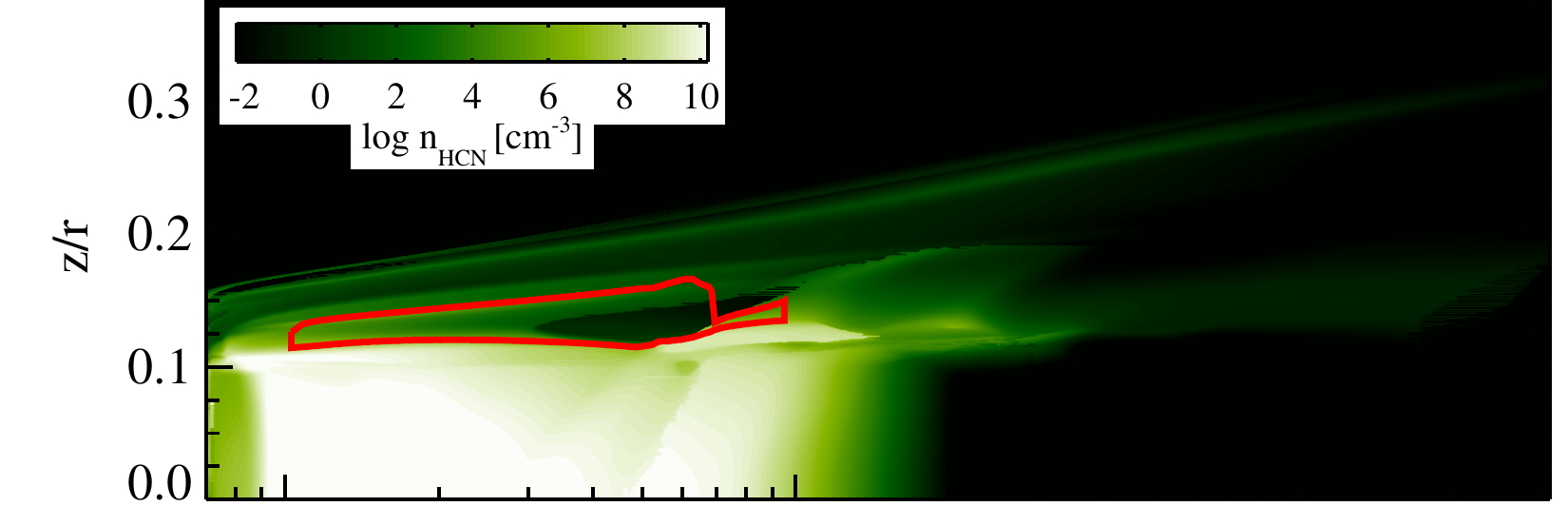}\\[-27mm]
{\color{white} HCN $\sf \lambda\!=\!14.0264\,\mu$m}
\hspace*{15mm}\\[22.3mm]
%\includegraphics[height=27mm,width=95mm,
%                 trim=15 38 20 290,clip]{Line_NH3_H_1034.pdf}\\[-27mm]
%{\color{white}\sf NH$_{\sf 3}$ $\sf \lambda\!=\!10.3376\,\mu$m}
%\hspace*{15mm}\\[22.3mm]
\includegraphics[height=35mm,width=95mm,
                 trim=-10 0 -4 287,clip]{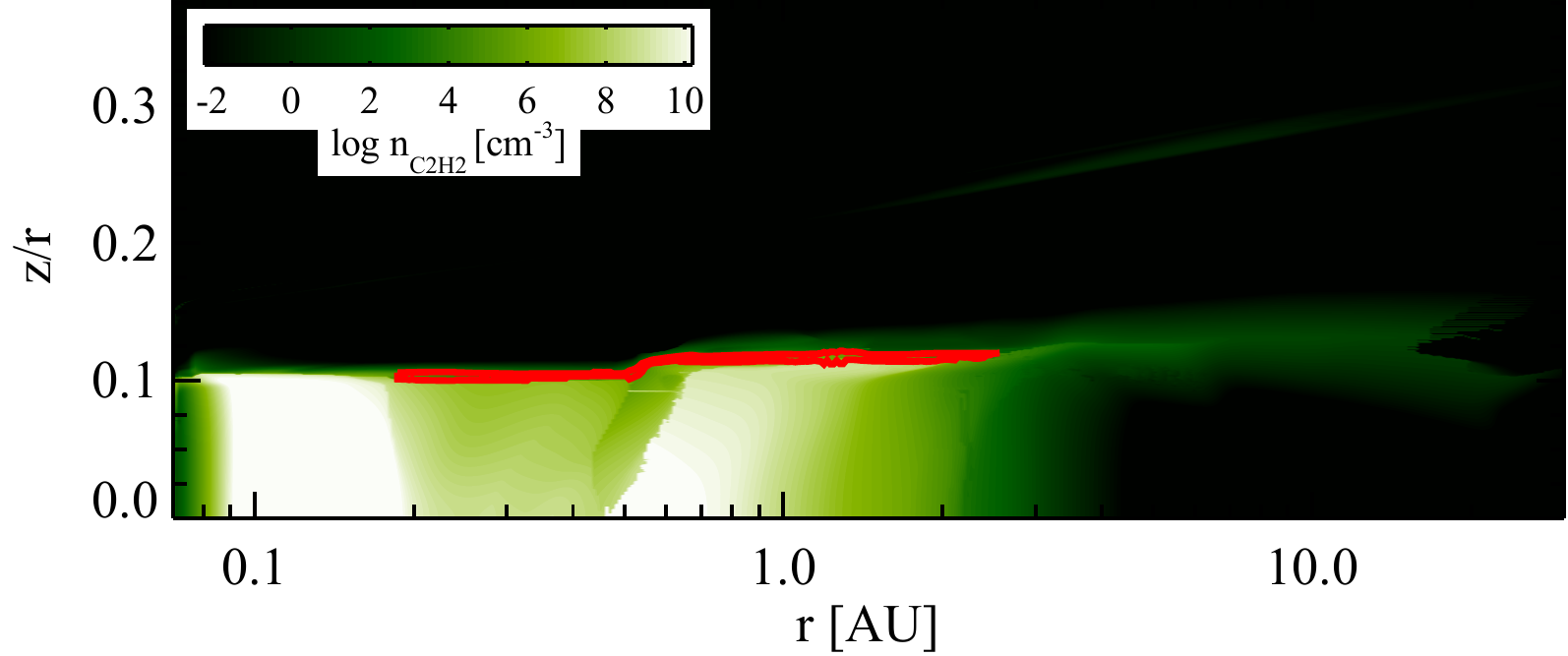}\\[-35mm]
{\color{white}\sf C$_{\sf 2}$H$_{\sf 2}$ $\sf \lambda\!=\!13.2039\,\mu$m}
\hspace*{15mm}\\[29.5mm]
\end{tabular}
\caption{Molecular particle densities and mid-IR line origin. The six
  contour plots show, from top to bottom, the calculated particle
  densities of OH, H$_2$O, CO$_2$, HCN, NH$_3$, and C$_2$H$_2$ in the
  main model. For each molecule, we selected a strong line and
  encircled the disc regions with a red line that are responsible for
  50\% of the fluxes of those lines.}
\label{fig:emregions}
\vspace*{-3mm}
\end{figure}
% OH_H   497 ->  418 lam[mic]= 2.011506E+01 Eu[K]= 5.527E+03 Aul[1/s]= 5.032E+01 Ncr[cm^-2]= 1.971E+12        X1/2   0,              13->        X1/2   0,  RR 12.5ff   12
% oH2O    78 ->   52 lam[mic]= 2.385980E+01 Eu[K]= 2.892E+03 Aul[1/s]= 3.128E+01 Ncr[cm^-2]= 1.794E+12     0    0    0,     9    8    1->     0    0    0,     8    7    2
% CO2_H   91 ->   24 lam[mic]= 1.497765E+01 Eu[K]= 1.113E+03 Aul[1/s]= 1.542E+00 Ncr[cm^-2]= 1.191E+14        0 1 1 01,                ->        0 0 0 01,      Q 16e     
% HCN_H  105 ->   31 lam[mic]= 1.402640E+01 Eu[K]= 1.306E+03 Aul[1/s]= 2.027E+00 Ncr[cm^-2]= 1.303E+14         0 1 1 0,                ->         0 0 0 0,      Q 11e     
% NH3_H  691 ->   25 lam[mic]= 1.033756E+01 Eu[K]= 1.515E+03 Aul[1/s]= 1.264E+01 Ncr[cm^-2]= 6.243E+13       0 1 0 0 a,   3  3 0  a    ->       0 0 0 0 s,   3  3 0  s    
% C2H2_H 382 ->   33 lam[mic]= 1.320393E+01 Eu[K]= 1.313E+03 Aul[1/s]= 3.509E+00 Ncr[cm^-2]= 8.417E+13  0 0 0 0 1 1  u,                ->  0 0 0 0 0 0+ g,      R 11e     

The thick sections of the abundance graphs in Fig.~\ref{fig:vertical}
highlight the line emitting regions, i.e.\ the layers that are found
to be responsible for 70\% of the observable line emissions from those
molecules in that column. To evaluate these quantities, {we use
  the escape probability method (Appendix~\ref{AppA}) to determine 
  the emission region of each spectral line, and then average over all
  lines emitted in the mid-IR region of that molecule according to}
\begin{equation}
  \langle X\rangle = \sum\limits_{\rm lines} X\, F_{\rm line} \;\bigg{/}\;
                     \sum\limits_{\rm lines} F_{\rm line}
  \label{eq:mean} \ ,
\end{equation}
{where $X$ can be any quantity we are interested in; for
  example the vertical hydrogen column density at the upper or lower
  end of the line forming region. The values $F_{\rm line}$ are the integrated
  fluxes of all mid-IR emission lines of a particular molecule, where
  we have used all mid-IR emission lines listed in Table~\ref{tab:IRdata},
  and all ro-vibrational lines around $4.7\,\mu$m for CO.
%all lines in $12-18\,\mu$m for H$_2$O, 
%CO$_2$, HCN and C$_2$H$_2$, and all lines around $26\,\mu$m for OH. 
  Table~\ref{tab:coldens} shows additional emission line flux averaged
  properties of the molecules; for example the mean continuum and line
  optical depths or the gas and dust temperatures in the
  line forming regions, using Eq.~(\ref{eq:mean}).}

\begin{figure*}
\vspace*{-2mm}
\hspace*{-3mm}
\centering
\begin{tabular}{cc}
\includegraphics[width=8.8cm,trim=30 27 120 85,clip]{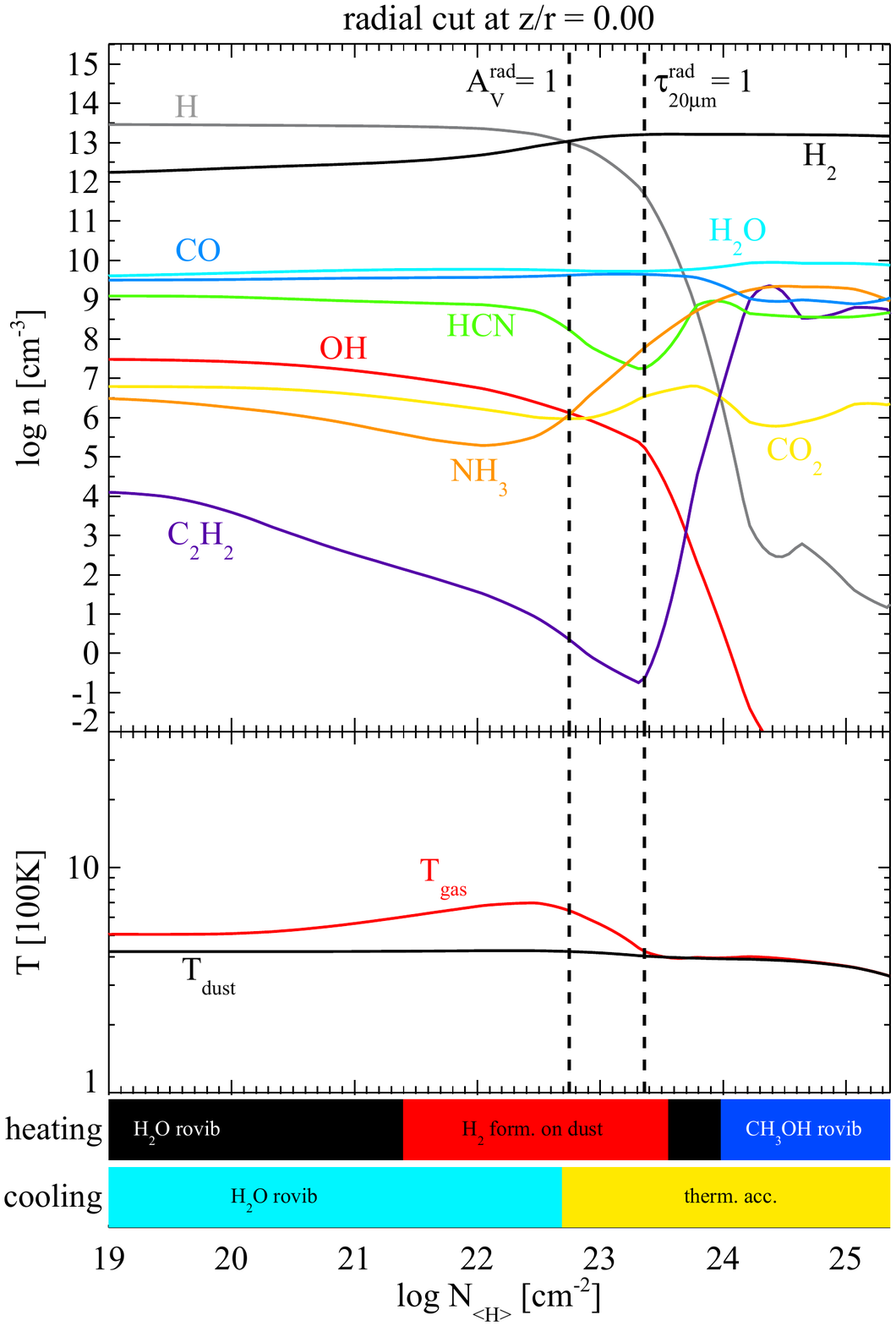}
\includegraphics[width=8.8cm,trim=30 27 120 85,clip]{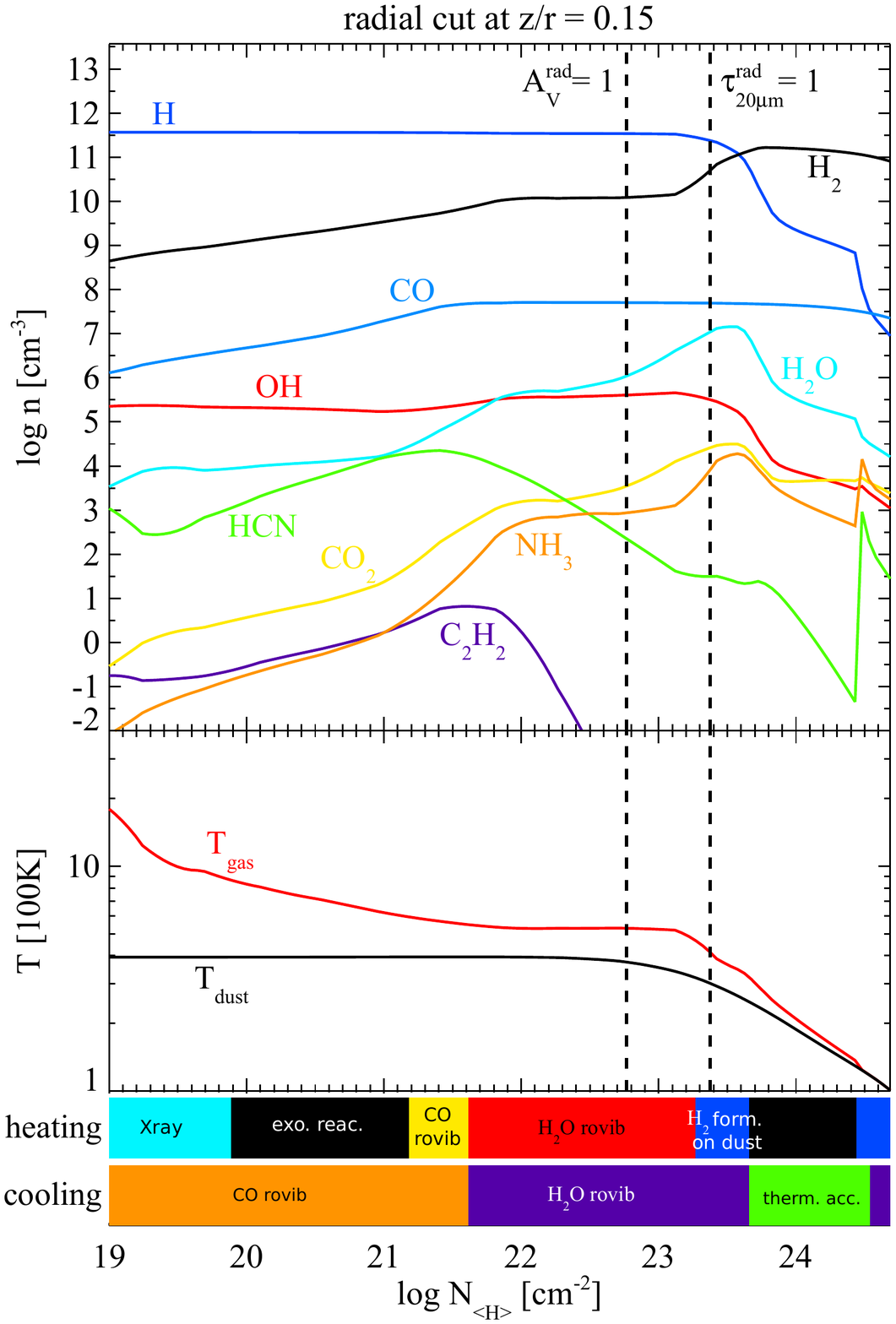}
\end{tabular}\\[-3mm]
\caption{Abundances of mid-IR active molecules along radial cuts
  at constant $z/r$ through the disc model with inner disc radius
  $R_{\rm in}\!=\!1\,$au and $\rm g/d\!=\!100$, plotted as
  a function of the radial hydrogen nuclei column density $N_\H$ (see
  text for how $N_\H$ corresponds to $r$). The
  two vertical dashed lines indicate where the radial dust optical
  depth is one ($A_{\rm V}^{\rm rad}=1$), and where the radial dust
  optical depth at $\lambda=20\,\mu$m is one. The lower plot shows the
  gas and dust temperature along that cut. The coloured
  bars below indicate the most important heating and cooling
  processes.}
\label{fig:radial}
\vspace*{-2mm}
\end{figure*}

The upper edges of the line forming layers shown in
Fig.~\ref{fig:vertical} usually coincide with a strong increase of
molecular concentration, and the lower edges with the line ($+$
continuum) optical depth becoming huge. This highlights an important
general finding of this work.  The observable lines of different
molecules probe the gas temperature in different layers of the disc,
just where these molecules start to become abundant, along the
following depth sequence:\ \ OH
(highest)\ \ ---\ \ CO\ \ ---\ \ H$_2$O and
CO$_2$\ \ ---\ \ HCN\ \ ---\ \ C$_2$H$_2$ (deepest).\ \ This is once
more illustrated in Fig.~\ref{fig:emregions}, where we plot the 2D
molecular concentrations and show how thin the line forming layers are
for single ro-vibrational lines. Resolving these line formation
regions spatially presents a numerical challenge to the model.

For additional orientation, in Fig.~\ref{fig:vertical} we indicate
the height at which the vertical dust optical depth at
$\lambda\!=\!20\,\mu$m is one, $\tau^{\rm ver}_{\rm 20\,\mu
  m}\!=\!1$. The C$_2$H$_2$ abundance does eventually reach high
values, but only below this height, so the C$_2$H$_2$ lines are
covered by dust continuum in our model. Figures~\ref{fig:vertical} and
\ref{fig:emregions} show that the disc model has no water ice in the
midplane at $r\!=\!0.3\,$au, whereas water ice is present at $r\!=\!1\,$au, as
evident from the missing gaseous H$_2$O. This creates a local
carbon-rich environment with gas element abundances $\rm C/O\!>\!1$, where
C$_2$H$_2$ is about two orders of magnitude more abundant in the
midplane than at $r\!=\!0.3$\,au.

The lower part of Fig.~\ref{fig:vertical} shows a few details about
the gas energy balance in the inner disc regions. We generally find
high gas temperatures of the order of several 1000\,K in the diluted,
photodissociated and partly ionised gas high above the disc. Once the
first molecules form (in particular OH and CO), their ro-vibrational
line cooling causes the gas temperature to drop to several 100\,K,
which leads to further molecule formation and accelerated
cooling. This atomic $\to$ molecular transition hence occurs very
suddenly and takes place well above the height $A_V^{\rm rad}\!=\!1$
at which the disc casts a shadow and the dust temperature starts to
drop. At the $A_V^{\rm rad}\!=\!1$ height, small dust particles scatter
the stellar UV photons partly downwards, which then penetrate deeper
into the disc until the vertical dust optical depth reaches about
$A_V^{\rm ver}\!\approx 3-5$. This effect generally leads to a
positive temperature contrast between gas and dust of the order of a
few 100\,K in the layers responsible for most of the observable line
emission.

There is an intermittent maximum of $T_{\rm gas}(z)$ as a function of
column densities around $\log_{10}N_\H\,{\rm[cm^{-2}]}\!\approx\!23.5$ in
Fig.~\ref{fig:vertical}, which is caused by optical depth effects in
the major cooling lines. At $A_V^{\rm rad}\!=\!1$ the line optical
depths are still small and molecular line cooling works very
efficiently. At slightly deeper layers, however, there is still some
heating by UV photodissociation followed by exothermal reactions and
H$_2$ re-formation on grain surfaces. But the line optical depths are
already large here, making the cooling ineffective, and the temperature
rises again. Eventually, the UV is completely absorbed and gas and dust
temperatures equilibrate through inelastic collisions (thermal
accommodation).  Figure~\ref{fig:vertical} also shows that a major part
of the line forming region, from $A_V^{\rm rad}\!=\!1$ to $\tau_{\rm
  20\,\mu m}^{\rm ver}\!=\!1$, is H rich. Thus, to discuss
the non-LTE population of ro-vibrational states in discs, we need collision
rates with atomic hydrogen.

In Fig.~\ref{fig:radial} we show two radial cuts
(along constant $z/r$) through the model with inner radius
$R_{\rm in}\!=\!1\,$au and $\rm g/d\!=\!100$. This time, the results
are depicted as a function of {\sl radial}\ \ hydrogen nuclei density
$N_\H$ as measured from the inner wall. The gas densities are very
high in this wall, $n_\H\!\sim\!10^{13.5}\rm\,cm^{-3}$ at $z/r\!=\!0$
and $n_\H\!\sim\!10^{11.5}\rm\,cm^{-3}$ at $z/r\!=\!0.15$. In order to
resolve the line formation in these walls, we need a radial grid with
initial increments of column densities or order
$\sim\!10^{19}\rm\,cm^{-2}$. This requires radial inter-point
distances as small as $\Delta r\sim\!10^5$\,cm close to the wall,
i.e.\ about 1\,km or $10^{-8}$\,au. The end of the line
emitting regions ($\tau^{\rm rad}_{20\rm\,\mu m}\!=\!1$) is reached at
about $N_\H\!=\!10^{\,23.5}\rm\,cm^{-2}$ which, in the model with
$R_{\rm in}\!=\!1\,$au, translates to $r\!=\!1.06\,$au at
$z/r\!=\!0.15$ and $r\!=\!1.0006\,$au at $z/r\!=\!0$. Close to these
walls, {\sc ProDiMo} automatically switches to a PDR description where
the molecular shielding factors, line escape, and pumping
probabilities are measured in the radial direction.

These results show, however, that the walls are in fact not PDRs. At
the high densities present in the wall, the two-body gas-phase
chemical rates dominate over the UV-photon rates even if the wall is
fully irradiated by the star, as in this model. {Consequently, the
  molecular concentrations in Fig.~\ref{fig:radial} are already high
  on the left side, where the wall starts.} The situation resembles
the case of thermochemical equilibrium, where the UV, however, still
plays a role in heating the gas.  For very high densities, close to
the midplane, we find radial layers in the wall where mid-IR lines of
one particular molecule (such as H$_2$O) simultaneously provide the
most important heating and cooling mechanism; that is the gas is in
radiative equilibrium in consideration of the water line opacity,
similar to for example brown dwarf atmospheres.

\subsection{Impact of chemical rate networks and C/O ratio}
\label{sec:chemistry}

\begin{figure}
\centering
\vspace*{0mm}
\hspace*{-3mm}\includegraphics[width=92mm]{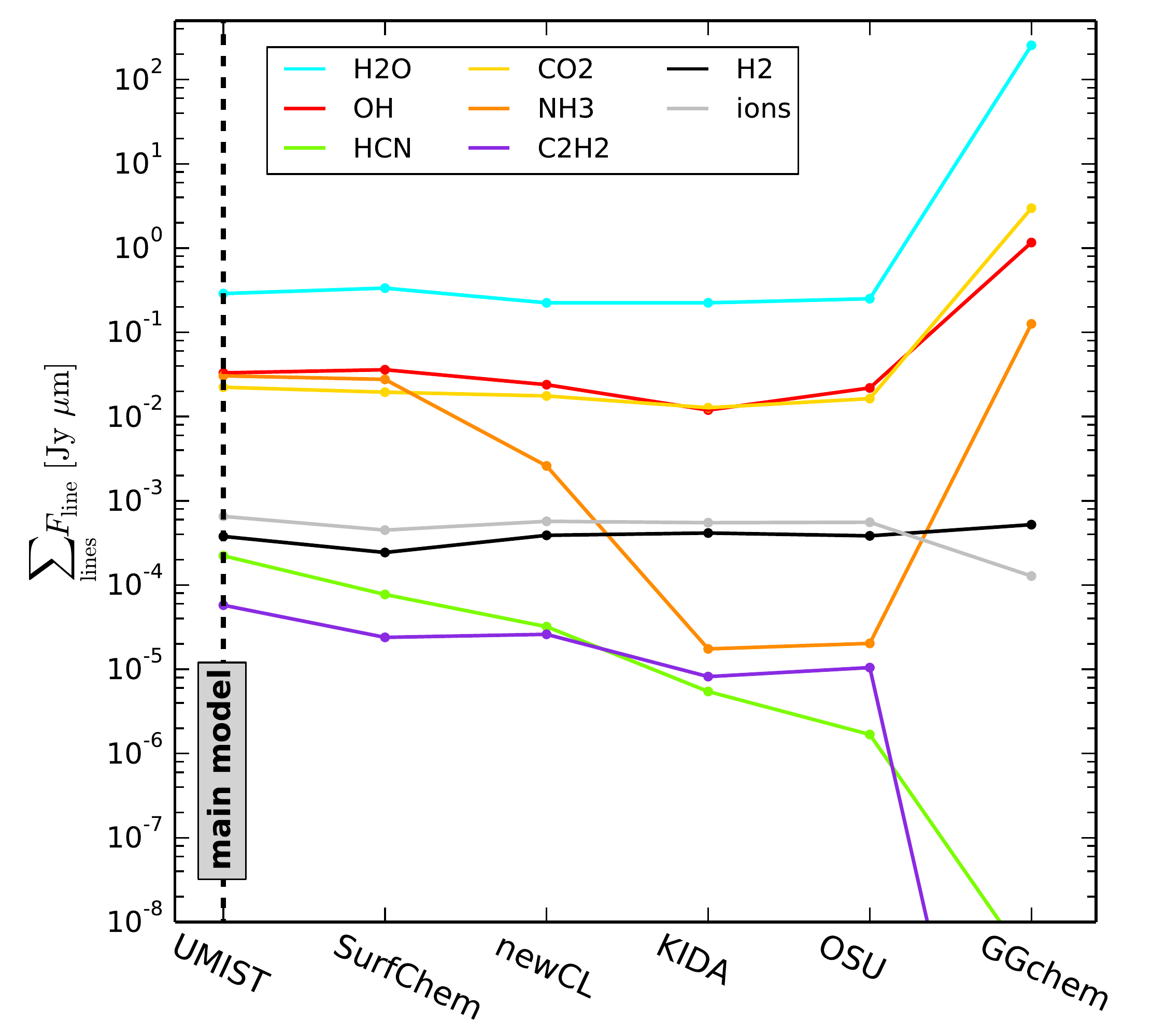}\\*[-2mm]
\caption{Total mid-IR molecular line emissions as a function of chemical rate
  network used in the model; see text for explanations.}
\label{fig:chemistry}
\vspace*{-2mm}
\end{figure}

The influence of the choice of the chemical rate network on the mid-IR
line flux results is studied in Fig.~\ref{fig:chemistry}.  The base
network used in this paper is UMIST~2012 \citep{McElroy2013} with
additional three-body (collider) reactions from UMIST 2006
\citep{Woodall2007}. All reactions among our selected species are
taken into account from this database, completed by simple ice
adsorption and desorption reactions, along with a few other special
rates for excited hydrogen and PAHs; for details see
\citep{Kamp2017}. The integrated mid-IR line fluxes of the molecules
and ions in this {\sf UMIST} model are again plotted on the left side
of Fig.~\ref{fig:chemistry}, denoted by ``main model''.  We then
re-computed this model with different base chemical rate networks
(using the gas temperature structure of the main model) as follows:
\begin{itemize}
\setlength{\itemsep}{3.5pt}
\setlength{\parskip}{0pt}
\setlength{\topsep}{0pt}
\setlength{\parsep}{0pt}
\setlength{\partopsep}{0pt}
\item {\sf SurfChem} refers to a new warm surface chemical model 
  (Thi et al.~2018 submitted).
\item {\sf newCL} is identical to the main {\sf UMIST} model, but has
  one three-body reaction rate reduced\footnote{The original reference
    \citep{Avramenko1966} states a rate constant of $10^{-32}~{\rm
      cm}^6~{\rm s}^{-1}$ with N$_2$ being the colliding partner in
    that experimental work. For reasons that we cannot trace, the
    value reported in the NIST database is six orders of magnitude
    higher, which is the value also found in the UMIST~2006
    database. KIDA~2014 does not include this reaction.
    \cite{Avramenko1966} stated that the bi-molecular rate constant is
    negligible compared to the three-body rate. Similar conclusions
    have been drawn from recent theoretical work by
    \citet{Galvao2015}.  The formation of NH$_2$ involves spin-orbit
    coupling and is highly forbidden, leading to a very low rate
    constant. Given that there are no more recent works on assessing
    this key three-body rate constant, we decided to be conservative
    and assign a rate constant that is of similar order of magnitude
    compared to all other collider reactions in UMIST~2006,
    $10^{-30}$~cm$^6$/s.}, for $\rm\, N \,+\, H_2 \,+\, M
  \,\longrightarrow\, NH_2 \,+\, M$, from $10^{-26}\rm\,cm^6/s$
  \citep{Woodall2007} to $10^{-30}\rm\,cm^6/s$.
\item {\sf KIDA} 2014 \citep[Kinetic Database for
  Astrochemistry,][]{Wakelam2012,Wakelam2013}.
\item {\sf OSU} 2009 (Ohio State University chemical network) from Eric Herbst.
\item {\sf GGchem} \citep{Woitke2018} is a fast thermochemical equilibrium
  code that computes all molecular concentrations based on the
  principle to minimise the system Gibbs free energy.
\end{itemize} 
We sorted the results according to the total HCN line flux, which
shows a variation of about two orders of magnitude (not counting {\sf
  GGchem}). Nitrogen is found to be predominantly atomic in the disc
surface layers according to the {\sf KIDA} and {\sf OSU} models, and
this is because these networks do not include the reaction\ \ $\rm N
\,+\, H_2 \,+\, M \,\longrightarrow\, NH_2 \,+\, M$.  The {\sf newCL}
results demonstrate the importance of that reaction, which ``activates''
the nitrogen chemistry and eventually leads to the production of
CN, HCN, and N$_2$ by follow-up neutral-neutral reactions. With reduced
collider rate coefficient, the {\sf newCL} results resemble the
{\sf KIDA} and {\sf OSU} results more.  \citet{Walsh2015} have also shown
the necessity to include three-body reactions into their network to model
the dense and warm chemistry in the planet-forming regions of
protoplanetary discs.

The line fluxes based on the {\sf GGchem} equilibrium chemistry model 
are extremely strong for water, CO$_2$, and OH. In thermochemical
equilibrium, there are much more molecules present in the surface of
the disc because photodissociation and other X-ray induced
destruction mechanisms are not considered. However, since the gas is
assumed to be oxygen rich, there are practically no HCN and C$_2$H$_2$
molecules in this model; therefore the emission features of those molecules
are strongly suppressed.  The bottom line is that none of the other
chemical rate networks used in combination with our disc model bring
us closer to the {\sc Spitzer/IRS} line observations, which would
require to increase the HCN and C$_2$H$_2$ lines fluxes while
decreasing those of CO$_2$.

\begin{figure}
\centering
\vspace*{-2mm}
\hspace*{-3mm}\includegraphics[width=92mm]{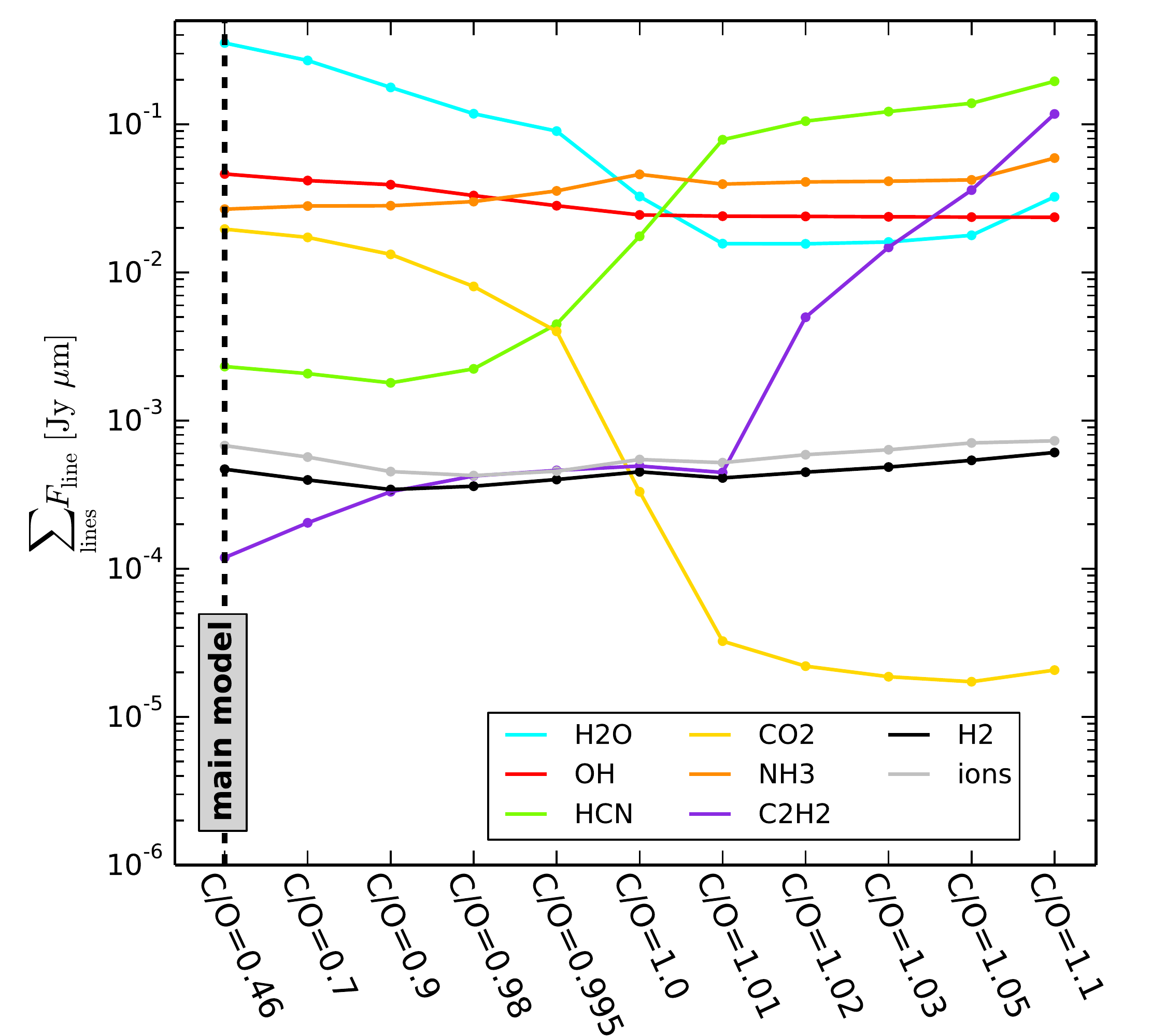}\\*[-1mm]
\caption{Total mid-IR molecular line emissions as a function of the
         carbon to oxygen ratio (C/O).}
\label{fig:CzuO}
\vspace*{-2mm}
\end{figure}

Another option is to consider an enrichment of the gas in the disc
surface with carbon at au distances, see results plotted in
Fig.~\ref{fig:CzuO}. An increase of the carbon to oxygen ratio (C/O)
$\to\!1$ actually shows the desired trend: the HCN lines become
stronger while those of CO$_2$ get weaker.  However, we do not
consider this option to be a fully viable solution either. The water lines
get too weak as C/O approaches unity, and when the C$_2$H$_2$ lines
finally become observable, the HCN and NH$_3$ lines are already 
much too strong. However, it is noteworthy that a modest increase of
C/O helps our model to get closer to the Spitzer observations.
\citet{Najita2013} discussed the HCN/water ratio as a function of C/O,
arriving at similar conclusions, and \citet{Pascucci2009,Pascucci2013}
related these results to the Spitzer data.

%============================================================================
\section{Summary and discussion}
%============================================================================

Previous analyses of the mid-IR molecular emission spectra of T\,Tauri
stars have mostly been based on parameterised modelling approaches
in which the temperatures and molecular concentrations are free
parameters, which are then fitted to the observed line emission
spectra. In the most simple case, single-point LTE slab models
\citep{Carr2011,Salyk2011} have been applied. More ambitious
single-point non-LTE slab and simplified 2D disc models have been used
by, for example \citet{Bruderer2015} and \citet{Bosman2017}, focussing on HCN
and CO$_2$, respectively. In an attempt to eliminate the current
uncertainties in chemical rate networks and heating/cooling physics in
discs, the scientific procedure in these works can be summarised as
follows: (i) derive temperatures and molecular column densities from
the line observations, (ii) divide the column densities by each other
to determine the molecular abundance ratios, and (iii) adjust element
abundances and chemical rate parameters in the models until agreement
is achieved.

In this paper, we have followed a very different strategy. We
used the full 2D chemical and temperature results from complex
thermochemical disc models. Without any detailed fitting, we 
showed that our model spectra are broadly consistent with the observed
properties of {\sc Spitzer/IRS} line emission spectra when we either
assume large gas/dust ratios or when we consider discs with directly
irradiated vertical walls at au distances.  A number of averaged
molecular properties from our models are listed in
Table~\ref{tab:coldens} for completeness.

\begin{figure}
  \vspace*{-1.5mm}
  \includegraphics[width=8cm,trim=100 50 130 90,clip]{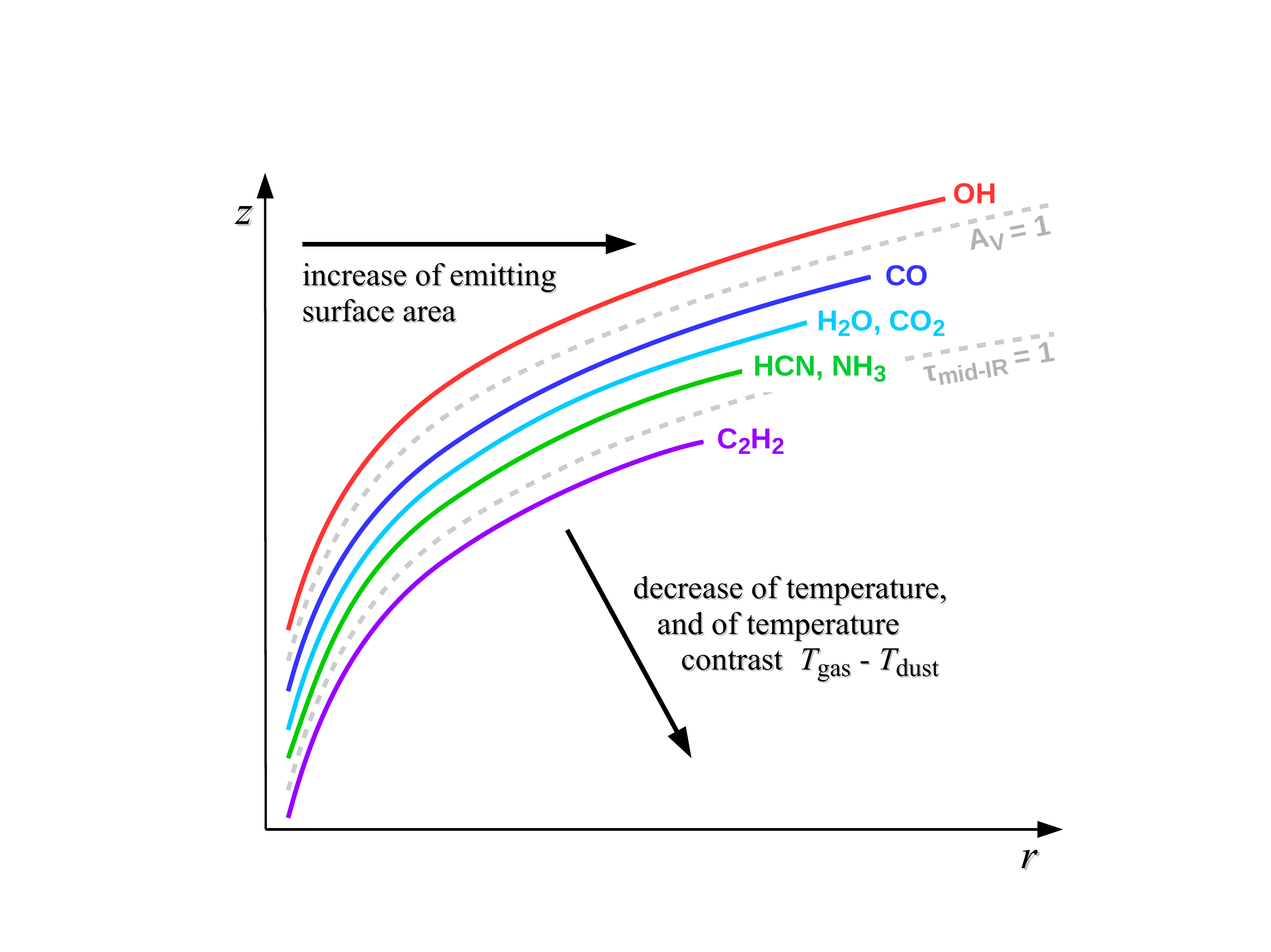}\\[-5.5mm]
  \caption{Onion-like shells of mid-IR line forming regions probing
    the disc temperature at different depths.}
  \label{onions}
  \vspace*{-0.5mm}
\end{figure}

We think that both modelling strategies are valid approaches, however,
we would like to highlight a few general results from our
modelling approach that we think are robust:\\*[-5mm]
\begin{itemize}
\setlength{\itemsep}{3.5pt}
\setlength{\parskip}{0pt}
\setlength{\topsep}{0pt}
\setlength{\parsep}{0pt}
\setlength{\partopsep}{0pt}
\item The mid-IR molecular lines from protoplanetary discs are
  optically thick and form above an optically thick dust continuum,
  therefore the temperature contrast between gas and dust
  has a decisive influence on the resulting line fluxes.

\item In PDR modelling, as for example performed in this paper, the
  molecular concentrations usually increase by many orders of
  magnitude with increasing depth.  As both density and concentration
  increase outside-in in discs, the mid-IR molecular line optical
  depths will reach unity in some layer. Most of the observed line
  photons from that molecule originate from this layer.  Subsequently,
  the line optical depths become huge and/or the continuum optical
  depths exceed unity. In both cases, the line flux contributions of
  the molecules situated in those deeper layers is small.

\item Consequently, the mid-IR lines of different molecules form in
  rather thin shells at different geometrical depths, where the
  physical conditions can be very different. We find the molecular
  lines in our models to be emitted from a succession of onion-like
  shells as sketched in Fig.~\ref{onions}, first OH, then CO, then
  H$_2$O etc., which we consider as a natural and straightforward
  result of the applied PDR physics.

\item Molecular column densities derived from observations can only
  provide averaged information about the concentration of the
  molecules above the $\tau_{\rm dust}\!=\!1$ disc surface, and these
  values physically depend on the dust opacities assumed. Therefore, a
  realistic treatment of the dust continuum is an important ingredient
  for any mid-IR line modelling.

\item Different lines of different molecules are emitted from
  different radial disc zones. Therefore, dividing column densities
  derived from observations by each other can produce misleading
  mixing ratios.

\item Studying molecular emission lines from the tenuous disc surface 
  alone might be incomplete as the molecular lines are also partly emitted
  from the inner rim where the line excitation and chemical
  conditions are different.
\end{itemize}

\subsection{LTE or non-LTE\,?}

\citet{Bruderer2015} and \citet{Bosman2017} have presented detailed
investigations of non-LTE effects in discs, including the pumping of
the ro-vibrational states by IR dust emission, concerning HCN and
CO$_2$, respectively.  Their conclusion is that non-LTE effects are
important for both flux and line shape, in particular with regard to
IR molecular bands at shorter wavelengths (3\,$\mu$m and 4.5\,$\mu$m
for HCN and CO$_2$, respectively).  \citet{Thi2013} presented a
thorough discussion of non-LTE effects on fundamental CO emission from
discs at similar wavelengths.

In contrast to these results, \citet{Meijerink2009}, using
parameterised chemical concentrations, and
\citet{Antonellini2015,Antonellini2016}, using full thermochemical
models, found no dramatic non-LTE effects for the water emission.  Our
results show no significant deviations of the water line fluxes
either, if we force the water molecules to be populated in
LTE. Concerning CO$_2$ and HCN, our models currently do not allow for
a non-LTE treatment.
%but preliminary studies with guessed collision rates also seem to
%indicate that the difference between LTE and non-LTE models are quite
%small. 
However, the ro-vibrational OH lines in our model, which are treated
in LTE, blend in nicely with the pure rotational OH lines, which are
treated in non-LTE. This agrees well with the TW\,Hya observations
(Fig.~\ref{fig:commonlines}).

We interpret this dispute about the importance of non-LTE effects as
follows. Our models suggest that the concentrations of most molecules
are vanishingly small in the upper disc layers and only become
suddenly abundant at rather deep layers, where gas densities are
already as large as $10^{12}\rm\,cm^{-3}$; see
Figs.~\ref{fig:vertical}.  Under such conditions, non-LTE effects are
expected to be small. In contrast, when assuming constant
concentrations in the entire column as in \citep{Meijerink2009},
\citep{Bruderer2015} and \citep{Bosman2017}, the lines are expected to
form at higher altitudes, i.e.\ in a tenuous environment where non-LTE
effects can be important. If the molecular lines are directly emitted
from vertical disc walls, wherein the densities are even higher, non-LTE
effects are expected to be even less significant.

However, more detailed investigations are required here. The
ro-vibrational lines of the OH radical are a particularly interesting
case because OH already forms at high altitudes where non-LTE effects
are likely to be important.

\begin{table*}
\caption{Mean molecular properties in the main disc model with $\rm
    gas/dust\!=\!1000$, in consideration of two vertical cuts;
    cf.\ Fig.~\ref{fig:vertical}.}
\label{tab:coldens}
\vspace*{-2mm}\hspace*{-2mm}
\resizebox{18.3cm}{!}{
\def\z{\!\!}
\begin{tabular}{c|c|c|c|c|c|c|c}
\hline
&&&&&&&\\[-2.2ex]
\z molecule\z & \z vertical cut\z 
         & $\langle\lambda\rangle\rm\,[\mu m]^{(1)}$
         & col.~density\,$[\rm cm^{-2}]^{(2)}$ 
         & $\langle$line formation$\rangle\,[\rm cm^{-2}]^{(1,3)}$
         & $\langle\tau_{\rm line}\rangle^{(1,2)}$
         & $\langle\tau_{\rm dust}\rangle^{(1,2)}$ 
         & $\langle T_{\rm gas}\rangle\rm\,[K]^{(1,3)}$\z\z \\
\hline 
&&&&&&&\\[-2.0ex]
OH     & $r\!=\!0.3$\,au 
     & 24.8 & $1.5(+17)$ & $4(+14)-2(+16)$ & 36        & 790 & 1180 \\
CO     & $r\!=\!0.3$\,au
     & 4.71 & $1.3(+23)$ & $1(+18)-4(+19)$ & $1.1(+7)$ & 1100 & 470 \\
CO$_2$ & $r\!=\!0.3$\,au 
     & 14.9 & $1.4(+21)$ & $6(+15)-8(+16)$ & $4.6(+5)$ & 840 & 480 \\
H$_2$O & $r\!=\!0.3$\,au
     & 18.7 & $1.3(+23)$ & $8(+17)-3(+19)$ & $1.4(+6)$ & 840 & 480 \\
NH$_3$ & $r\!=\!0.3$\,au
     & 11.3 & $2.2(+22)$ & $6(+15)-1(+17)$ & $2.8(+6)$ & 970 & 460 \\
H$_2$  & $r\!=\!0.3$\,au
     & 15.9 & $5.0(+26)$ & $4(+22)-8(+23)$ & 810       & 840 & 360 \\
HCN    & $r\!=\!0.3$\,au
     & 14.0 & $5.2(+21)$ & $9(+13)-1(+16)$ & $1.9(+6)$ & 860 & 330 \\ 
C$_2$H$_2$ & $r\!=\!0.3$\,au
     & 13.8 & $8.2(+19)$ & $3(+14)-1(+17)$ & $1.8(+4)$ & 860 & 230 \\
\hline
& & & & & & &\\[-2.2ex]
OH     & $r\!=\!1$\,au
     & 24.3 & $5.9(+16)$ & $2(+14)-1(+16)$ & 12        & 230 & 720 \\
CO     & $r\!=\!1$\,au
     & 4.69 & $9.9(+20)$ & $2(+16)-7(+18)$ & $1.8(+5)$ & 340 & 360 \\
CO$_2$ & $r\!=\!1$\,au
     & 15.0 & $8.8(+16)$ & $3(+15)-2(+16)$ & 31        & 250 & 330 \\
H$_2$O & $r\!=\!1$\,au
     & 22.2 & $1.4(+19)$ & $2(+17)-4(+18)$ & $1.2(+3)$ & 250 & 320 \\
NH$_3$ & $r\!=\!1$\,au
     & 11.1 & $4.4(+17)$ & $3(+15)-3(+16)$ & 90        & 300 & 320 \\
H$_2$  & $r\!=\!1$\,au
     & 17.1 & $1.5(+26)$ & $2(+22)-2(+23)$ & 330       & 250 & 280 \\   
HCN    & $r\!=\!1$\,au
     & 14.1 & $6.1(+20)$ & $5(+14)-3(+17)$ & $3.2(+5)$ & 260 & 140 \\   
C$_2$H$_2$ & $r\!=\!1$\,au
     & 13.8 & $1.2(+21)$ & $5(+16)-7(+18)$ & $6.4(+5)$ & 260 & 130 \\
\hline
\end{tabular}}\\[1mm]
\hspace*{1mm}\begin{minipage}{18.1cm}
\footnotesize 
$^{(1)}$: Mean properties denoted by $\langle\cdot\rangle$ are
    calculated according to Eq.\,(\ref{eq:mean}) as average over all
    included mid-IR emission lines listed in
    Table~\ref{tab:IRdata}, weighted by line flux,\ \ 
$^{(2)}$: Maximum values reached in the midplane,\ \ 
$^{(3)}$: Values in the line forming regions.
\end{minipage}
\end{table*}

\subsection{Surface or wall emission?}
\label{sec:walls}

Detecting disc gaps at au scales associated with planet formation is a
difficult task, even with the currently best observational
techniques such as IR interferometry or ALMA
\citep{ALMA2015,Pinte2016}. The information obtained from continuum
data is necessarily limited to the dust component, but does not reveal
the structure of the gas. 

As this paper demonstrates, mid-IR molecular emission lines are partly
emitted from the disc surface and partly emitted from irradiated disc
walls (Fig.~\ref{FLiTs-ProDiMo1}). We have only discussed truncated
discs with fully irradiated walls in this paper\footnote{It would
  obviously be very difficult to explain the accretion phenomenon in
  such discs when there is no gas at all close to the star up to au
  distances.} and further modelling of discs with gaps is certainly
required. But we can say with confidence already that there are
significant spectroscopic differences (strength, colour, and ratios of
the emission features of different molecules) between disc surface
emission and disc wall emission (see Fig.~\ref{fig:escpro}). These
differences are caused by the different temperature, density, and
radiation field conditions in the disc surfaces compared to those in
the walls. Therefore, high S/N observations of spectroscopically
unresolved mid-IR molecular emission spectra, as will become possible
with JWST, might offer an alternative way to detect and characterise
disc walls at au distances in T\,Tauri stars.

Considering discs with gaps, only a small fraction of the vertical
walls at the outer end of a gap might be fully irradiated, or these
walls might be completely submerged in the shadow of a tall inner
disc, in which case the implied changes of the mid-IR line spectrum
are expected to be small. But if there are directly irradiated parts
of a disc wall present at au scales, it might be possible to detect these with
JWST based on the spectroscopic fingerprints of molecular wall
emission. Our models are well suited for subsequent research on this
matter as the models cover all relevant physics and chemistry.

\section{Conclusions and outlook}

We have used full 2D thermochemical disc models to calculate complete
atomic and molecular emission spectra from T\,Tauri stars between
$9.7\,\mu$m and $38\,\mu$m, using a mixture of LTE and non-LTE
techniques.  We introduced {\sc FLiTs} to ray trace tens of thousands
of mid-IR molecular emission lines in one shot.

Without detailed fitting of disc shape or dust opacities, we find a
reasonable agreement with previously published {\sc Spitzer/IRS}
spectra. To achieve this agreement, however, we need to increase the
gas/dust ratio from 100 to about 1000 at radial distances of a few au,
or consider transitional discs which have distant inner walls at a few
au or illuminated secondary walls following gaps as expected for
planet-disc interactions.

Generally speaking, strong mid-IR lines are emitted from exposed,
irradiated molecules. In our models, such molecules exist when (i) the
gas/dust ratio is large; (ii) the dust is unusually transparent in the
UV, for example when all small grains are removed; or when (iii) dust
settling is very strong.  Concerning the third option, the sub-micron
grains (the main UV opacity carriers) would need to be removed from
the disc surface by gravitational settling at au distances. According
to the physical description of dust settling used in this paper
\citep[][applying the midplane gas density]{Dubrulle1995}, we find it
difficult to explain the bright mid-IR line emission from T\,Tauri
stars by settling.  \citet{Antonellini2015,Antonellini2016} arrived at
similar conclusions.  However, \citet{Riols2018} have proposed an
improved treatment of settling, taking into account the decrease of
gas density with height, which fits their numerical mixing
experiments. According to Riols \& Lesur's settling description, it
seems indeed possible to remove the sub-micron grains from the disc
surface by gravitational settling. More numerical experiments are
required to verify this solution.

Another possibility is to have (iv) molecules situated in dense
illuminated vertical disc walls, where the two-body gas phase
reaction rates are huge and can compensate the losses by UV
dissociation. The resulting molecular concentrations are rather
independent of the UV irradiation, and thus favour the existence of
irradiated molecules.

In all these scenarios, large radiation fields overlap with large
molecular concentrations, and the heating UV photons are absorbed by
the molecules rather then by the dust.  More detailed investigations
are required to study to what extent we can disentangle wall emission
from disc surface emission, and to possibly use new JWST line
observations to detect discs with illuminated secondary walls at au
scales.  Taking into account complementary spectral energy
distribution and continuum IR visibility data should allow us to
reject some of these scenarios.

Future non-LTE investigations require collision rates with atomic
hydrogen because the line forming regions are partly H rich and
H$_2$ poor, as is true for CO \citep{Thi2013}. Ro-vibrational lines of
the OH radical are expected to show the strongest non-LTE effects
because these lines form at high altitudes above the disc.

In all our models that broadly agree with the observed mid-IR line
strengths, the mid-IR lines of H$_2$O, OH, CO$_2$, HCN, and C$_2$H$_2$ are
optically thick. Therefore, we conclude that mid-IR line fluxes are not
good tracers of column densities. Instead, we find that the gas/dust
temperature contrast has a decisive influence on the strength and
shape of the IR molecular emission spectra. We see no direct influence
of the ice lines on the emitted IR spectra, as ice formation takes
place only very deep in the disc midplane ($A_V\!\ga\!10$). However,
vertical mixing processes could establish a link to the ice lines.

The chemistry in the planet-forming and IR line emitting regions of
protoplanetary discs needs further attention, including {
  three-body reactions}, warm surface chemistry, and combustion. At the
moment, {models with different chemical rate networks predict
  rather different mid-IR spectra, and our standard chemical rate
  network seems to somewhat overpredict CO$_2$, and underpredict HCN
  and C$_2$H$_2$. These molecular concentrations are strongly affected
  by the assumed C/O ratio in the gas. For $\rm C/O\!\to\!1$, we would
  expect more HCN, more C$_2$H$_2$, and less CO$_2$ to form, which
  could help to better understand the smultaneous emissions from all
  these molecules other than water.  If oxygen is consumed in deeper
  layers to form water ice, there is a cold trap for oxygen in the
  midplane, and if some turbulent mixing establishes a physical link
  from the spectroscopically active surface layers to that cold trap,
  the oxygen abundance would actually be expected to decrease over
  time.}

\bigskip\noindent {\bf Acknowledgements:\ } {We thank the
  anonymous referee for the insightful remarks that helped
  improve the paper.} The research leading to these results has
received funding from the European Union Seventh Framework Programme
FP7-2011 under grant agreement no 284405. The computer simulations
were carried out on the UK MHD Consortium parallel computer at the
University of St Andrews, funded jointly by STFC and SRIF.

\bibliography{references}

\appendix

\section{Escape probability model}
\label{AppA}

\begin{figure*}
\vspace*{-2mm}
\centering
\hspace*{-3mm}
\includegraphics[width=18.8cm]{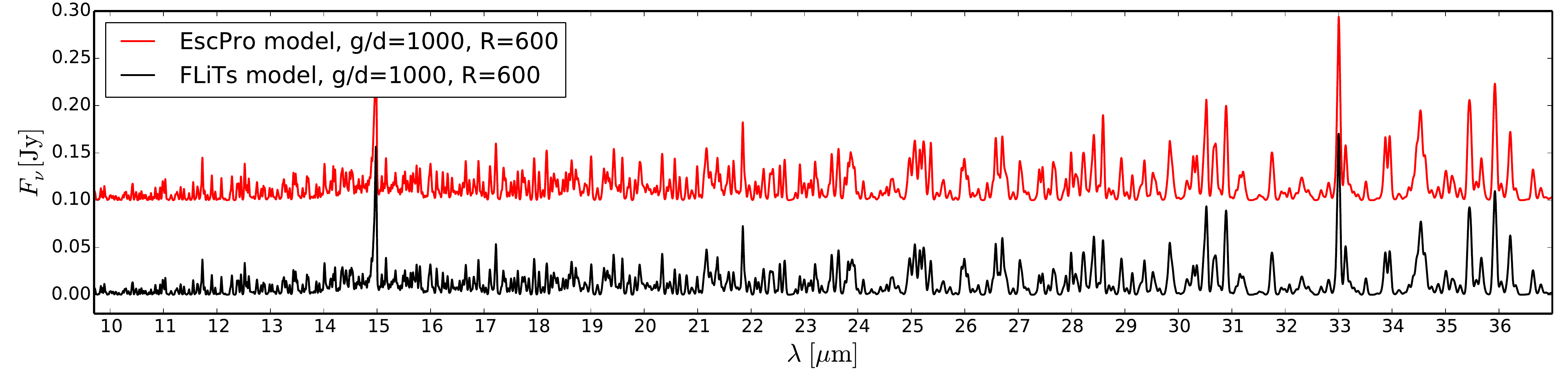}\\[-4.5mm]
\hspace*{-3mm}
\includegraphics[width=18.8cm]{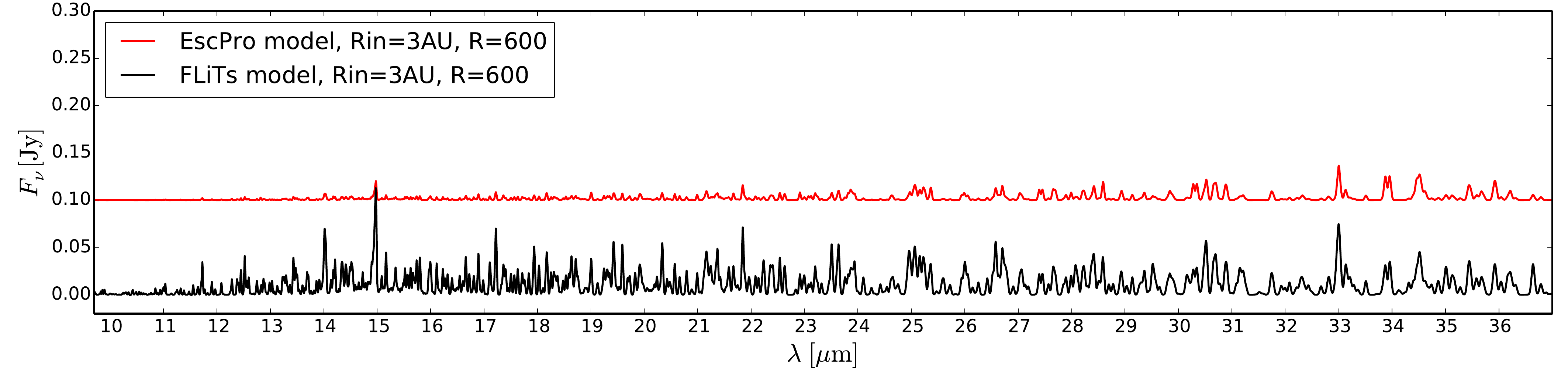}\\[-1mm]
\caption{Comparison between continuum-subtracted {\sc FLiTs} spectra
  (black), ray traced for disc inclination $45^\circ$, and spectra
  directly obtained from our vertical escape probability method (red,
  ``EscPro''), both convolved to $R\!=\!600$ spectral resolution. The
  agreement is remarkable for the main model with gas/dust ratio $\rm
  g/d\!=\!1000$ and inner disc radius $R_{\rm in}\!=\!0.07\,$au (upper
  plot), where disc surface emission dominates. However, for the
  truncated disc model with $\rm g/d\!=\!100$ and $R_{\rm
    in}\!=\!3\,$au (lower plot), the escape probability method fails because the
  line photons are mostly emitted sideways through the inner disc
  wall, in which case the EscPro spectrum results to be too faint, and
  too red.}
\label{fig:escpro}
\end{figure*}

The details of our simplified non-LTE treatment of atomic and
molecular level populations (escape probability method) are explained
in \citep[][Sect.~6.1 therein]{Woitke2009}. This article provides the
general concept, physical quantities and symbols used. Once the model
is completed, we can calculate the vertically escaping line luminosity
[erg/s] of transition $u\!\to\!l$ as
\begin{equation}
  L_{\rm line} = \sum\limits_{\rm cells} 
                n_u V_{\rm cell}\;h\nu_{ul}\;A_{ul}\; 
                P_{ul}^{\rm pump}\!(\tau_{ul}^{\rm ver})\; 
                \exp(-\tau^{\rm ver}_{\rm dust}) \ ,
  \label{Lcell}
\end{equation}
where $n_u$ [cm$^{-3}$] is the upper level population, $V_{\rm cell}$ [cm$^3$]
the volume of a computational cell, $\nu_{ul}$ the line centre frequency, and
$A_{ul}$ [s$^{-1}$] the Einstein emission coefficient. For the escape
probability, we take the probability of line photons to reach the disc
surface on a vertically upward pointed ray; see $P_{ul}^{\rm pump}$ in
Eq.~(83) of \citep{Woitke2009}. Vertical line centre and continuum
optical depths are given by
\begin{eqnarray}
  \tau_{ul}^{\rm ver} &=& \frac{A_{ul} c^3}{8\pi\nu_{ul}^3}
                \int_z^\infty \frac{1}{\Delta\nu_{\rm D}}
                \left(n_l\frac{g_u}{g_l}-n_u\right)\;dz'
  \label{eq:tauline}\\
  \tau^{\rm ver}_{\rm dust} &=& \int_z^\infty \kappa^{\rm ext}_{\nu_{ul}}\;dz'
  \label{eq:taudust} \ ,
\end{eqnarray}
where $\Delta\nu_{\rm D}$ is the (thermal$\,+\,$turbulent) Doppler
width of the line. To calculate Eq.~(\ref{eq:taudust}) we use a simple
Simpson integration based on the given dust particle densities and
opacities. Concerning Eq.~(\ref{eq:tauline}), we perform an implicit
integration by assuming that $n_u/n_l$ is constant in each cell and
that the line optical depth used to calculate the population numbers
refer to the bottom of each cell. In order to account for the lower
half of the disc, the disc is mirrored at the midplane and the optical
depths are computed across the disc to the other surface, such that
each computational cell actually occurs twice in Eq.~(\ref{Lcell}).
Finally, the line luminosity is converted into line flux at distance
$d$ as
\begin{equation}
  F_{\rm line} = \frac{1}{4\pi\,d^2}\,L_{\rm line} \ .
\end{equation}
Equation~\ref{Lcell} provides an easy way to identify the cells (or
spatial disc region) that are most important for the emission
of a given spectral line. We have used this concept to highlight
the line emission layers and regions in Figs.~\ref{fig:vertical},
\ref{fig:emregions}, and Table~\ref{tab:coldens}.

Figure~\ref{fig:escpro} shows that the agreement with the full
ray-tracing {\sc FLiTs} spectrum is excellent for the main model, whereas we
cannot apply the escape probability method to discs with au-sized
inner holes, where the lines are mostly emitted by the inner disc
wall. To summarise,
\begin{itemize}
\setlength{\itemsep}{1.3pt}
\setlength{\parskip}{0pt}
\setlength{\topsep}{0pt}
\setlength{\parsep}{0pt}
\setlength{\partopsep}{0pt}
\item The escape probability spectra often provide very good first
  guesses of the line emission spectra and come for free with {\sc ProDiMo},
  i.e.\ they do not need any additional computational time to be calculated. 
\item The escape probability technique always predicts line
  emission. This is physically not guaranteed. Infrared line
  absorption is expected for large disc inclinations and for active
  discs that are heated internally by accretion, where the vertical
  temperature gradients are reversed.
\item The escape probability spectra do not contain any
  line profile information, but are useful only for spectral 
  resolutions up to a couple of 1000.
\item The escape probability technique cannot be applied to
  transitional discs with inner disc radii larger than a few
  $1/10^{\rm th}$ of au, as the line photons in such discs rather
  escape sideways through the inner disc wall.
\end{itemize}

\end{document}